\documentclass[a4paper,fleqn]{cas-sc}

\usepackage[authoryear,longnamesfirst]{natbib}
\usepackage{lineno}

\def\tsc#1{\csdef{#1}{\textsc{\lowercase{#1}}\xspace}}
\tsc{WGM}
\tsc{QE}

\begin{document}
\let\WriteBookmarks\relax
\def\floatpagepagefraction{1}
\def\textpagefraction{.001}

\shorttitle{Sun Close-Encounter Model of Long-Period Comet and Meteoroid Orbit Stochastic Evolution}    

\shortauthors{Pilorz S., Jenniskens P.}  

\title [mode = title]{Sun Close-Encounter Model of Long-Period Comet and Meteoroid Orbit Stochastic Evolution}  



%

\author[1]{Stuart Pilorz}

\cormark[1]{}


\ead{spilorz@seti.org}


\credit{Conceptualization, Methodology, Numerical modeling, Writing and Editing}

\affiliation[1]{organization={SETI Institute},
            addressline={339 Bernardo Ave, Ste 200}, 
            city={Mountain View},
            postcode={94043}, 
            state={CA},
            country={USA}}

\author[1,2]{Peter Jenniskens}


\ead{pjenniskens@seti.org}


\credit{Conceptualization, Data Collection, Writing and Editing}

\affiliation[2]{organization={NASA Ames Research Center},
            addressline={Mail Stop 244-11}, 
            city={Moffett Field},
            postcode={94035}, 
            state={CA},
            country={USA}}

\cortext[1]{Corresponding author}



\begin{abstract}
The dynamical evolution of long-period comets (LPCs) and their meteoroid streams is usually described with the Sun as the primary body, but over most of their orbits the Solar System barycenter (SSB) is effectively the orbital focus. Detailed numerical integrations show that the orbital elements in the barycentric reference frame are nearly constant, except within the orbit of Jupiter where the comet or meteoroid shifts to a heliocentric orbit. Here we show that this encounter can be modeled in the barycentric frame analogously to how planetary close encounters are treated in the heliocentric frame, with the comet captured into an elliptic orbit about the Sun as it in turn orbits SSB. Modeling the encounters as a two-body interaction in the SSB frame gives a different insight into the dynamics than offered by secular perturbation analyses, and reveals that a large portion of the stochasticity seen in the evolution of the comet’s orbit is due to the Sun’s state relative to SSB at the time of encounter. LPCs sample the Sun's state randomly at each return, so that a statistical characterization of Sun's state is sufficient to determine the qualitative evolution of their orbits, including stream dispersion. The barycentric orbital elements are shown to execute random walks well-characterized by Maxwellian distributions. This is superimposed atop a systematic orbital precession induced by planetary torques. Planetary close encounters add a second stochastic component, but this component does not typically dominate the solar perturbations. Based on the statistics of Sun’s state alone, the age of a long-period comet meteoroid stream in a given orbit can be derived to reasonable precision from the observed random dispersion of the angular orbital elements at Earth.
\end{abstract}

\begin{highlights}
\item 
A long period comet mostly orbits the solar system barycenter except inside Jupiter's orbit where it orbits the Sun.
\item 
Perturbations to a long period comet's orbit are explained by a two-body close-encounter with the Sun as it orbits the solar system barycenter.
\item
 The long period evolution is a Markovian random walk resulting from random sampling of the Sun's state at encounter.
\item
 Meteoroid streams disperse predominantly from stochastic changes in energy and angular momentum experienced during Sun close encounter
\end{highlights}


\maketitle

\section{Introduction}\label{}

More than half of the roughly 500 known meteor showers at Earth are from Long period comets (LPCs) that pass close to Earth's orbit \citep{Jenniskens2023}. These comets are Potentially Hazardous Objects and a source of meteoroids that are a danger to spacecraft. Only LPCs with orbital periods less than 4,000 years have dense enough meteoroid streams to be detected at Earth as meteor showers \citep{Jenniskens2021}. The dispersion of those streams at Earth informs about how long ago a comet ejected those meteoroids.

\begin{figure}[p!]
\begin{center}
\includegraphics[width=7in]{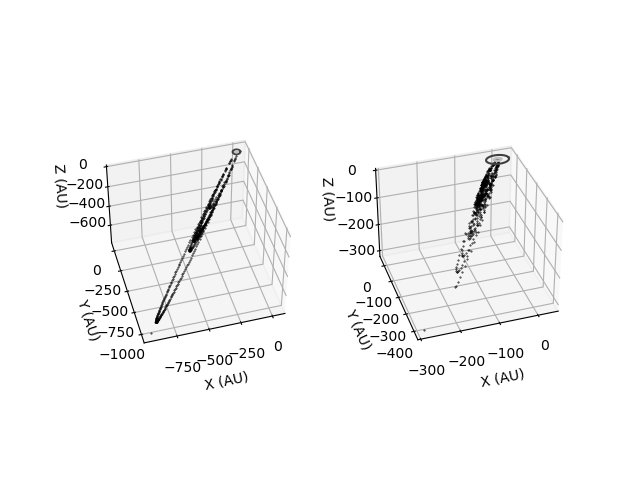}
\caption[]{
  \label{Fig1_Orbital_Stream}
  The far reaching orbits of LPCs compared to the orbits of Neptune, Saturn, and Jupiter (top right). Individual points are for a 60kyr evolution shower for a comet with median elements of shower \#206, starting with orbital periods of 4000 years (left) and 250 years (right). 
	}
\end{center}
\end{figure}

LPCs have orbital periods in excess of 200 years, and return to the inner solar system at a time when Jupiter and Saturn, which predominantly influence them, are at positions unrelated to those during the previous return \citep{Chambers1997}. By definition, LPCs have orbits that are not prone to settling into strong resonances with those planets.
Figure~\ref{Fig1_Orbital_Stream} illustrates just how far LPCs travel beyond the giant planets during most of their orbit. The figure shows the result from numerical simulations of 60kyr evolution of an LPC orbit with nominal periods of 4000 years (left) and 250 years (right).  Neptune and Jupiter's orbits are shown in both panels as dark and light gray circles, respectively.

In the absence of mean-motion resonances and planetary close encounters, the orbit of an LPC, and those of its ejected meteoroids, is expected to disperse randomly from its original orbit over time, while the nodal line and line of apsides of the orbit gradually change due to precession. The random part of the dispersion correlates with the age of the stream, typically determined on a case-by-case basis from numerical models. 

Most of the long term study of LPCs and associated meteor streams is conducted numerically.  Numerical models are fast and accurate enough to accomplish all the tasks of directly modeling known bodies.  They allow researchers to follow streams for long periods of time, or search for suitable initial conditions when looking for matches between parent bodies and known streams (e.g., \citep{Hajdukova2019, Hajdukova2021a, Hajdukova2021b, Neslusan21}). These models show that meteoroid streams disperse over time, but do not explain why.  In this study, we back away from modeling individual bodies, and instead use numerical integrations to probe the phenomena driving LPCs.

The secular evolution of LPCs and their meteoroid streams is dominated by the gravitational interaction with the Sun and planets. In addition to gravitational forces, comets experience non-gravitational forces from gas and dust ejection, while meteoroids are affected by radiation and solar wind pressure, as well as radiation driven forces such as the Poynting-Robertson and Yarkovsky effect. These effects, as well as general relativity, are generally included in the numerical modeling of orbital evolution of comets and their meteoroids, but do not dominate the orbital evolution of LPCs or large meteoroids.

From an observational point of view, the stochastic component of an LPC's orbital evolution includes a nodal wandering in and out of Earth's orbit, a phenomenon that makes encounters with recently ejected dust of LPCs episodic in nature. \citet{Jenniskens1997} first showed that this wandering is well described by the Sun's reflex motion around the Solar System barycenter, primarily a consequence of Jupiter's 12 year orbital period and Saturn's 30 year period around the Sun, causing a 60-year recurrent pattern of shower sightings. Numerical modeling by \citet{Lyytinen2003} confirmed that cometary dust trails wander in and out of Earth's orbit in sync with the Sun's reflex motion.  They found that perturbations on the orbital period of a meteoroid are strong enough that sections of a comet dust trail catch up on each other after just one revolution, leading to an initial rapid dispersion. After that, the dispersion of LPC meteoroid streams is more gradual, but until now it was not understood why that is.

\citet{Jenniskens2023} found simple relations between the dispersions in the measured angular osculating elements and stream age. Dispersions in the inclination, $i$, argument of perihelion, $\omega,$ and node, $\Omega$, appear to grow linearly in time in such a way that their root-sum-square seems to increase linearly with age. Individual simulations showed some variation in the rate, but it was unclear how it depended on the orbital elements. Recently, \citet{Jenniskens2023} and \citet{Pilorz2023} have investigated the growth of dispersion over time in meteor streams associated with LPCs, with the aim of constraining the ages of observed streams.  These studies noted that the dispersion growth depends on the number of orbits completed rather than the time elapsed, and that the dispersion growth rate is inversely proportional to the perihelion distance.

Here, we expand on this work with the aim of understanding the physics behind these inferred relations.  \citet{Carusi1987} noticed that the orbits of Halley-type comets shift from barycentric to heliocentric focus.  We find this to be true of LPCs as well.  Unlike previous studies, we investigate the exchange of energy and momentum during the solar capture from the viewpoint of a single close-encounter between two bodies both orbiting the solar system barycenter (SSB), and show that the LPC period changes due to gravitational boost or braking by the Sun.  The Sun is not generally in the comet's orbital plane as it approaches barycenter, so that the capture also results in a change of the comet's orbital plane and subsequent release into yet a different one.  This alters the angular momentum direction. Torque from the planets is present and important in forcing the nodal and apsidal precession, and perturbations from close encounters with planets happen with some regularity, but even so it appears that the state of the Sun during the LPC's transit of the inner solar system determines a large part of the stochastic variation observed in the energy and angular elements of the LPC. 

The paper is organized as follows. The next section 2 describes the numerical methods used to calculate a comet's orbital evolution. In Section 3, we describe the secular evolution of orbits, and elucidate the role of Sun in changing the orbital elements. In section 4, we develop a toy model to study the statistics of perturbations resulting from the Sun's location alone and the expected dependencies on the orbital elements of the comet orbit.  In section 5, we cast the orbital evolution in the framework of a Markovian random walk, and apply the model to the rate of dispersion growth in the orbital elements of LPCs and their meteoroids. Section 6 provides a summary of our conclusions.

\section{Methods}

We use a combination of numerical integration and ephemeris-based Monte Carlo calculations to investigate the behavior of comets as they pass though the inner solar system.  

Long term (up to 100 ka) numerical simulations were performed on 29 established long-period comet meteoroid streams (Table~\ref{table1}) that are described in Chapters 4 and 5 of \citet{Jenniskens2023}. These streams sample different orbital configurations and were chosen to cover the phase space of inclination, $i,$ perihelion distance, $q$, and argument of periapsis, $\omega$. Four of those streams have known parent comets, including 6 LYR - April Lyrids from comet C/1861 G1 (Thatcher), 206 AUR - Aurigids from C/1911 N1 (Kiess), and 145 ELY - eta Lyrids from C/1983 H1 (IRAS-Araki-Alock). Simulations were performed of clouds of hundreds of particles dispersed around the orbit of the comet or the median orbital elements of the stream, starting from orbital periods of 250, 500 or 4000 years.  Twelve of these, whose numbers and names appear in boldface in Table~\ref{table1}, were selected for further study here, and a single orbit from each was followed in greater detail through many perihelion returns so that the stochastic wandering of the orbital elements could be investigated. Shower \#16 was run with three different values of inclination, and so appears in the Table as 16, 16a, and 16b.

The numerical code used here has been described in \citep{Jenniskens2023}. The core of the code is the 15th order quasi-implicit scheme from \citet{Everhart1969}, adapted for use with comets and meteoroids by \citet{Vaubaillon2005}.  We have made various improvements for increasing speed while retaining accuracy, and have altered the step control and output.

Our version of the code can use planetary and cometary ephemerides from either NAIF/SPICE \citep{Acton1996} or Calceph \citep{Gastneau2024}.  It is configured to run individual comets or clouds of ejecta using fixed or variable time steps, writing out the states at every node crossing.  In order to take advantage of the larger time steps that are allowed when away from perihelion, when running with a cloud of particles, a recursive method is used to group particles in true anomaly so that superfluous small time-step calculations are not spent on particles near aphelion. For the current study, in addition to tabulating all node crossings, fixed time steps are set at which all particle states are output. These are set for years to decades near aphelion, with the resolution increasing to 50 days inside 250AU, 25 days inside 50AU, and one day steps inside 6AU. This still does not give fine enough resolution to resolve events around each perihelion with the accuracy this study requires.  For that purpose we use the state vectors output from the longer runs near each perihelion as initial conditions to perform short, near-Sun integrations with time steps of a few hours.  The fine integration allows accurate evaluation of the times and states corresponding to perihelia.

Calculations are done by default in heliocentric ecliptic J2000 coordinates.  Validation has been performed comparing the integrator's results to NAIF comet ephemerides for short period comets.  The e-folding divergence from JPL results when run on the same IC's with the same kernels is approximately 500 years. For this application we do not integrate the planets and Sun, but use a +-500kyr ephemeris for gravitational forcing when integrating comet orbits \citep{Gastneau2024} .  Initial conditions (ICs) for long integrations of LPC orbits are found by integrating several perturbations on observed elements backwards for 100kyr, then using one of these that integrates forward to the present without getting perturbed to a period of >4000 or <250 years. We follow that comet from initial conditions from 100kyr ago to the present, recording the state at all node crossings and at time intervals as described above.  

At each output time step, we also tabulate the state of the Sun relative to the solar system barycenter from the ephemeris.  These and the comet states are read into a data structure that contains both the heliocentric ecliptic J2000 and barycentric ecliptic J2000 using the Sun's tabulated state.  The corresponding Keplerian osculating elements, $a, q, e, i, \omega, \Omega,$ and $M,$ are calculated in the usual way from the state vectors in both reference frames, and stored in the data structure.

For purposes of analyzing the Sun's perturbations, we calculate the distance above or below the orbital plane, $x_{\odot,\perp},$ as the inner product of the barycentric Sun position vector with unit vector in the direction of the barycentric angular momentum.  We also compute the Sun's azimuthal position in the ecliptic XY plane using polar coordinates, $(r_\odot,\phi_\odot)$, with $\phi_\odot$ measured prograde from the incoming barycentric longitude of ascending node at the initial orbit. Finally, for investigations of planetary forcing, our pipeline allows us to re-access the ephemeris to pull out planetary states at selected tabulated output times.  To quantify the comets' reactions to  solar state, we use Monte Carlo simulations with random solar locations pulled from the ephemerides. 

For purposes of comparing the calculated dispersions to observations, we take  the angle at which Earth traverses the stream into consideration.  While there is a gradual broadening of the stream over time, precession has the effect of making a shower visible on Earth for only part of the comet orbital evolution, resulting in changes of orbital elements along Earth's path.

\section{Results}

We describe below the general behavior of secular evolution seen in the integrations, present evidence that the comet is captured into a heliocentric orbit near periapsis, then focus on aspects from the integrations that show how the solar encounter affects the orbital energy and angular momentum, relating these to changes in the orbital elements.

\subsection{General Behavior of Orbital Evolution}

\subsubsection{Sequences of barycentric orbits}

It has been noticed that orbits of Halley-type comets transition between barycentric and heliocentric foci \citep{Carusi1987}.  We find this to be true of LPCs as well.  Most obviously, across all showers studied here, and for all orbital elements,  the barycentric orbital elements remain fixed and stable for most of each comet orbit (or, in the case of mean anomaly, change at constant rate) while the heliocentric elements oscillate due to the moving reference frame.  The barycentric elements change from orbit to orbit, as seen in the spread in semimajor axis of the along-orbit points shown in Figure~\ref{Fig1_Orbital_Stream}.

\begin{figure}[p!]
\begin{center}
\includegraphics[width=7in]{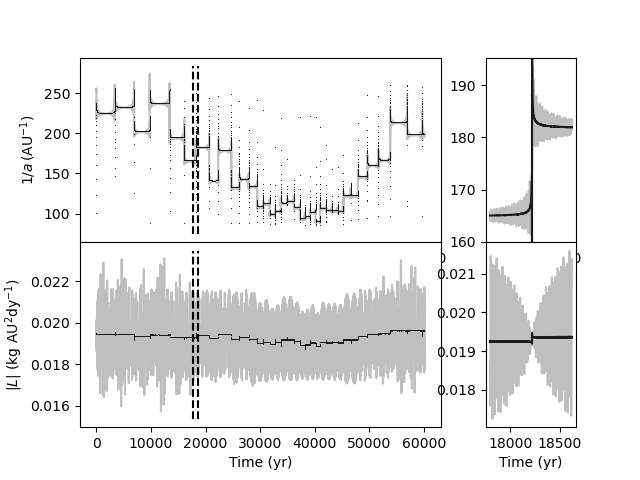}
\caption[]{
	\label{Fig3_Quantum_EL}
Evolution of the energy (expressed in terms of $1/a$) and the magnitude of the angular momentum in the direction of the ecliptic pole, $L$, during $60 {\rm kyr}$ of evolution for a comet with the median elements of shower \#647. The panels on the right are an expansion of the range in time bracket by dashed lines in the left hand panel. The elements in heliocentric coordinates are in light-gray, while the barycentric elements are in black.  For most of the duration of the orbit, between perihelia, the barycentric elements remain roughly constant with the heliocentric elements oscillating about them.  Both sets of elements undergo abrupt transitions at each perihelion.
}
\end{center}
\end{figure}

As an example, Figure~\ref{Fig3_Quantum_EL} shows the energy (top) and angular momentum (bottom) over 60kyr of a comet orbit based on the elements of shower \#647. Results in barycentric coordinates are in black, while those in heliocentric coordinates in light gray. The barycentric orbits take on steady values during most of each orbit, separated by brief transitions at perihelia (appearing as spikes in the figure).  In contrast, the elements calculated in heliocentric coordinates oscillate, due to the Sun's motion around the barycenter in an epicyclic motion reflecting the position and mass of the planets.  

The Sun's motion is characterized by the 12 year period of Jupiter and the 30 year period of Saturn, as well as the 60 year recurrent cycle of Jupiter and Saturn's orbital periods.  The amplitude of that motion is of order $0.010 AU$, just larger than the diameter of the Sun itself ($0.0093 AU$). About $0.005 AU$ of that displacement is due to Jupiter, another $0.003 AU$ from Saturn, with Uranus and Neptune accounting for much of the remaining $0.002 AU$ \citep{Jenniskens1997}.  The connection between semimajor axis and period, $P  \propto a^{3/2}$, can be seen in Figure~\ref{Fig3_Quantum_EL}, in that orbits with smaller semimajor axes (large $1/a$) have shorter periods and correspond to smaller segments on the time axis.

The right-hand panel of Figure~\ref{Fig3_Quantum_EL} zooms in on the perihelion that is bracketed by dashed lines in the left-hand panels to show the transition in more detail. It shows a typical case in which the energy and angular momentum of the orbit in barycentric coordinates transition from one value to another near perihelion over a period of a few hundred days.  

In Figure~\ref{Fig3_Quantum_EL}, the amplitude of the oscillations in the heliocentric energy is seen to increase when approaching perihelion, whereas the oscillation amplitude in angular momentum decreases.  This is due to the fact that at perihelion, the magnitude of the comet's radial vector becomes small, so that the Sun's motion perturbs it by up to 1\%.  At the same time, the comet's velocity becomes large enough that the Sun's perturbation to it is on the order of one part in $10^7$, so that $v_{\rm helio}\sim v_{\rm  SSB}.$  While the incoming comet is still effectively orbiting SSB, as it nears the Sun the effect of the radial perturbations due to the Sun's position and speed relative to SSB  cause the heliocentric energy and angular momentum direction to oscillate.  Conversely, the magnitude of the angular momentum is dominated by the velocity near perihelion, and the oscillations in the heliocentric elements related to angular momentum become less apparent at perihelion.

We find that the transitions in orbit are abrupt for all orbit configurations and for all the elements, except for the nodal angle, which is affected by the torque from the planets (as well as the Sun).  It undergoes a more gradual transition, which commences as far as 10AU from perigee, and which will be discussed in more detail below.

\subsubsection{Nature of the Transitions}

\begin{figure}[p!]
\begin{center}
\includegraphics[width=5in]{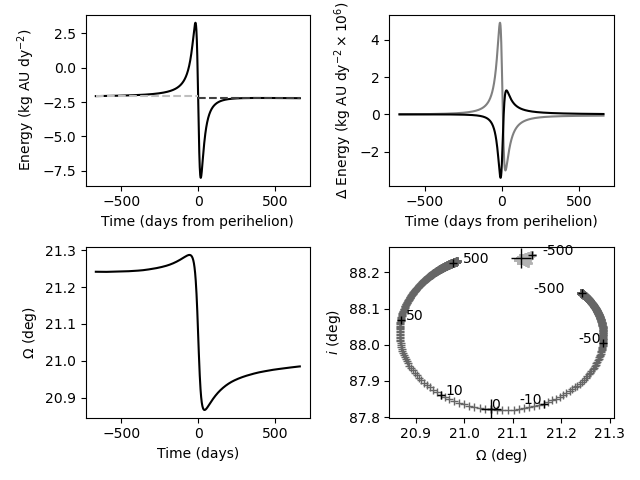}
\caption[]{
	\label{Fig4_merge_dEdL}
    Changes in total energy (top) and angular momentum (bottom) during a single perihelion passage of shower \#1055. Top-left shows the total energy calculated in the barycentric frame, and the top-right hand panel shows the the difference between the actual kinetic energy computed during the encounter from what it would have been had the orbital elements remained fixed at their pre-encounter values (light-gray) and the analogous difference in potential energy (dark-gray).  The differences are computed as the putative barycentric value minus the simulation value. The bottom left-hand panel shows the angle of the ascending node, $\Omega$, versus time, and the bottom right-hand panel shows the variation of the associated direction of angular momentum in the $i-\Omega$ plane during the same time period for the barycentric (dark-gray) and heliocentric (light-gray) elements.  Along the barycentric trajectory, the points corresponding to 500, 50, and 10 days from perihelion are labeled, and in both coordinate systems the point at perihelion is marked with a cross.  
}
\end{center}
\end{figure}

The transitions of the orbital elements at perihelion contain information about the physics behind them.
Figure~\ref{Fig4_merge_dEdL} shows in more detail the transitions in energy and angular momentum across an arbitrary perihelion.
The top two panels of Figure~\ref{Fig4_merge_dEdL} detail the change in the barycentric energy. The left-hand panel shows the energy versus time, as it rises above its initial value, then drops precipitously to a low value, before gradually rising to the value it will maintain for the duration of the next orbit.  This would appear as one of the spikes in the lower resolution depiction of Figure~\ref{Fig3_Quantum_EL}.  These spikes, visible at all transitions in Figure~\ref{Fig3_Quantum_EL}, are not numerical artifacts, but occur because of asymmetries in the trade off between kinetic and potential energy about SSB as the comet is tugged by the Sun.  

The top-right-hand panel of Figure~\ref{Fig4_merge_dEdL} shows the differences between the kinetic (light gray) and potential (black) energies of the numerically integrated orbit versus what the energies the comet would have had, had it remained on its incoming orbit and not been perturbed near perihelion.  Those putative energies, ${\rm KE}_{\rm SSB,in}$ and ${\rm PE}_{\rm SSB,in}$, were estimated by using the orbital elements calculated 600 days before perihelion, assuming they remained fixed, and extrapolating the state to subsequent output times merely by changing the true anomaly.

The sharp rise and drop in the energy seen in the left-hand panel result from the comet having been tugged by the Sun so that it has a higher and then lower kinetic energy than it would have been expected to have at the same radius on the initial barycentric orbit.  

In this case, a net deceleration occurs.  On approach to perihelion, momentum exchange with the Sun accelerates the comet relative to the barycentric orbit it would have followed. The total energy rises so that the comet appears less tightly bound in the barycentric frame. But at perihelion, the comet donates momentum to the Sun, and the kinetic energy becomes lower than what would be expected were the comet orbiting the barycenter at that radius.  Consequently the total energy rises suddenly. As the comet climbs out of the Sun's potential well and starts to feel the rest of the solar system, its velocity excess drops so that the velocity asymptotically approaches a value that balances the potential energy, but in this case after losing energy (gravitational braking), to emerge with a lower total energy than before the interaction.

The ways in which the potential and kinetic energy play catch-up vary with the orbital geometry. There are configurations among our simulated comets where the comet's velocity is initially increased by the Sun before decelerating into perihelion, and other cases where the velocity profile peaks symmetrically with the potential energy, but more sharply peaked. These result in falling-then-rising profiles for kinetic energy, instead.

A similarly abrupt profile is seen in the changes in the barycentric angular momentum direction during 600 days around perihelion (bottom panels of Figure~\ref{Fig4_merge_dEdL}).
The bottom left-hand panel shows the temporal change in nodal angle, $\Omega,$ during a transition at perihelion, where it has excursions above the incoming value and below the final outgoing value. The bottom right-hand panel shows the change in direction of the angular momentum vector during a perihelion crossing, plotting the inclination, $i,$ versus $\Omega$ during these 500 days on either side of perihelion, tabulated in one-day intervals.  The pair of angles, $(i,\Omega),$ specifies the direction of the angular momentum vector, expressed in spherical coordinates as a declination (angle from the ecliptic pole) and an azimuth (direction in-plane). The declination value measured from the Z axis is equivalent to the value of $i$, which is measured as an inclination from the X-Y plane.  The azimuth of the node vector, $N\equiv Z\times L,$  is $90^\circ$ removed from the azimuthal angle of the angular momentum, $L,$ projected into the ecliptic.  

In the figure, the directions of the barycentric angular momentum during the perihelion crossing are shown in dark gray, and the heliocentric angular momentum direction is in light gray.  The points corresponding to 500, 50, and 10 days before and after perihelion are marked along the barycentric trajectory, and to make the temporal direction clear a star is placed at the point 500 days before.  The trajectory of the heliocentric angular momentum is smaller, and the day numbers are not shown.  On both trajectories, the points at perihelion are marked with a large black cross.  At perihelion, the heliocentric angular momentum settles into a nearly constant direction while the barycentric angular momentum vector precesses. The barycentric vector does not precess about the heliocentric one, for reasons discussed below, but travels along a circle that also contains the heliocentric vector's direction.  Simulations show that excepting when planetary close encounters occur, it does not complete this circle but exits perihelion crossing with a slightly different orientation than coming in. The short spikes in the value of $\Omega$ seen in the left-hand panel and Figure~\ref{Fig3_Quantum_EL} prior to and after perihelion are seen to be the part of the precession, which involves an oscillation in both $i$ and $\Omega$.

\begin{figure}[p!]
\begin{center}
\includegraphics[width=6.5in]{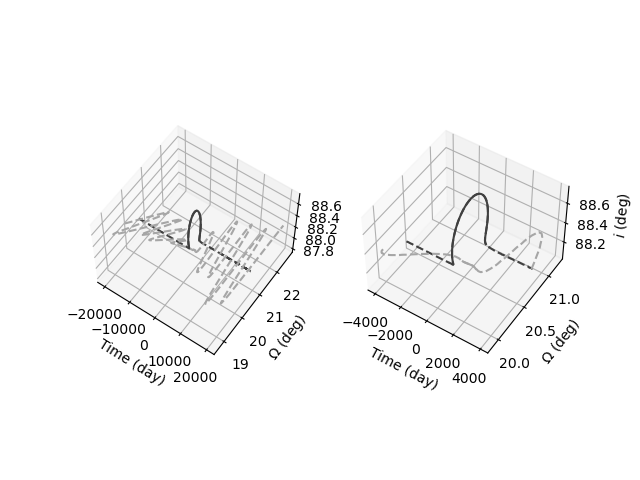}
\caption[]{
\label{Fig9_3D_Lvst}
3D plot of angular momentum changes in both heliocentric and barycentric coordinates. The change in the direction of the angular momentum vector, represented as $(i,\Omega)$ is shown versus time during a perihelion encounter from the median orbit of shower \#1055.  Time is labeled in days from perihelion and moves from the back to the front of the figure, inclination $i$ is plotted vertically and node $\Omega$ horizontally. The left-hand panel shows a period of 20,000 days on either side of perihelion, and the right-hand panel zooms in on the 4000 days on either side.  The barycentric angular momentum is traced in dark-gray, and the heliocentric in light-gray.  The curves are solid in the region where calculations of the type performed for Figure~\ref{Fig7_5AU_Transition} indicate that the comet is in orbit about the Sun, and are dashed elsewhere.
}
\end{center}
\end{figure}

Figure~\ref{Fig9_3D_Lvst} shows a 3-dimensional rendering of the precession of the angular momentum vector as it occurs over time, for a perihelion crossing from shower \#1055.  The direction of the angular momentum vectors calculated in the barycentric frame (dark gray) and heliocentric frame (light gray) are shown versus time, which is labeled in days from perihelion and moves from the back to the front of the figure.  The direction of the angular momentum vector is again characterized by $i$, plotted vertically, and $\Omega$, plotted horizontally. The left-hand panel shows 20,000 days either side of perihelion, and the right-hand panel zooms in to show only 4000 days on either side.  In both panels, the curves tracing the angular momentum directions are in dashed lines except for the times near perihelion when the force points at the Sun (as shown in Figure~\ref{Fig6_FLdr_vsAU}, and Figure~\ref{Fig10_TorqueOmega_vs_AU}, below), in which case they are solid.

Most obviously, the angular momentum with respect to the SSB (dark gray) is nearly constant on the incoming and outgoing orbits, but precesses near perihelion when the comet is in a heliocentric orbit.  The precession is fast and large, and the total value of $\Delta\Omega$ across perihelion depends upon how far through the precession the comet exits the Sun's influence.  The heliocentric angular momentum direction (light gray), exhibits gradually decreasing oscillations and then remains constant during the time when the Sun is in possession of the comet and the barycentric angular momentum is precessing.

The figure shows a simple case with no planetary close encounters.  Without planetary close encounters, the comet exits the Sun's sphere of influence fairly symmetrically with where it entered it, which is reflected in the precession: the barycentric angular momentum direction traverses a circle that also contains the heliocentric direction, and it starts and stops at fairly equal angular distances form the heliocentric direction, but on opposite sides.  The heliocentric orbital plane is of intermediate orientation between the incoming and exiting barycentric orbital planes, and generally equally inclined with respect to both.  If planetary encounters occur, they introduce a stochastic perturbation to the angular momentum direction either before or after the precession seen in the figure.

In this 3-dimensional graphic, we can see that what appears in Figure~\ref{Fig4_merge_dEdL} to be a linear drift in the heliocentric angular momentum direction as shown prior to encounter is actually a series of temporal oscillations in $i$ and $\Omega$.  At any time, the oscillations fall in along a narrow ellipse in the $(i,\Omega)$ plane that has a slowly changing slope, so that they appear planar in this figure.  The linear relation results from the fact that the Sun is approximately confined to the ecliptic X-Y plane.  The orientation of the line of oscillation depends on the angle between the Sun's orbital plane about SSB and the comet's, and so is different for the inbound and outbound barycentric orbits.  As noted in reference to Figure~\ref{Fig3_Quantum_EL}, the amplitude of the oscillations decreases with approach to perihelion because the percent perturbations to the barycentric velocity due to the Sun becomes small as the velocity of the comet increases.


\subsubsection{Framework for LPC orbit evolution}

Guided by these results, we proceed to investigate the orbital evolution in terms of an approximation where one stable barycentric orbit instantly changes into the next, due to forcing that happens briefly as the comet passes through the inner solar system.  All of our simulations can be succinctly characterized by sequentially specifying the elements of each barycentric orbit, and we use this framework below to investigate the perturbations of the elements between successive barycentric orbits.  We will find that a brief capture of the comet into a heliocentric orbit near periapsis explains the perturbations.

\begin{figure}[p!]
\begin{center}
\includegraphics[width=7in]{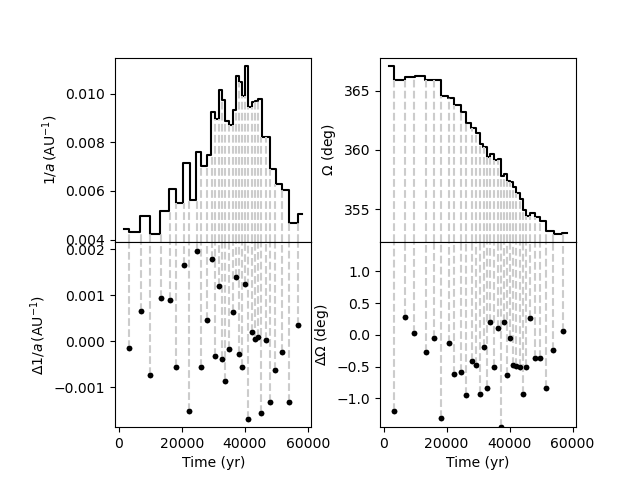}
\caption[]{
	\label{Fig5_Quantum_EL_wdeltas}
Changes in total energy (left) and angular momentum (right) during a full 60ka evolution. The upper panels show a discretized approximation in which for most of the orbit the elements remain constant at their aphelion values, approximating the behavior seen in Figure~\ref{Fig3_Quantum_EL}.  In the bottom panels the changes in the elements that occur at each perihelion are shown, connected by dashed lines with the perihelia they correspond to.  The secular evolution can be pictured as a series of orbits whose barycentric elements remain constant for the entire orbit;  dual to this is the sequence of changes in the elements at each perihelion.
}
\end{center}
\end{figure}

For each perihelion of each shower studied here, we compute the change in barycentric osculating element as the difference between its values at the succeeding and preceding aphelion, suitably averaged to avoid noise. 
Figure~\ref{Fig5_Quantum_EL_wdeltas} shows the time-wise progression of the barycentric energy (left) and nodal angle, $\Omega,$ (right) for a 60kyr evolution of shower \#647.   On the top on each side, we approximate the barycentric elements as being constant in time between successive perihelia at the value that they have near aphelion, but with 'instantaneous' jumps at perihelion.  On the bottom, connected by dashed lines to the transitions in the top plot, we show the accompanying change in the value that occurs at each perihelion.   

In the rest of the work below, we condense the description of orbits into the information as shown in Figure~\ref{Fig5_Quantum_EL_wdeltas}.  The evolution of an orbit will be summarized by citing the osculating elements at each aphelion, and, redundantly or alternatively, for each element, a perturbation,``$\Delta$'', at perihelion that changes it to the value it will hold across the next aphelion.

\subsection{The Sun Close-Encounter}

\subsubsection{Evidence for a heliocentric section}

Numerical simulations reveal three facts that point to the fact that the comet is captured into an elliptical orbit about the Sun during the period around peihelion when it transitions from one barycentric orbit to the next.  First, the angular momentum about the Sun becomes constant during that time; second, near perihelion, the direction of the force felt by the comet switches from SSB to the Sun; and finally, the integrated orbit is well approximated about perihelion as an elliptical orbit about the Sun.  


The behavior of the angular momentum can be seen clearly in Figure~\ref{Fig9_3D_Lvst}, above, which follows the angular momentum direction versus time for the barycentric and heliocentric frames, and is representative of all perihelion crossings studied here.  The solid portions of the trajectories highlight the timespan within a few hundred days of perihelion where the force vector points at the Sun.  During this time the angular momentum in the Sun's frame remains constant, while the angular momentum calculated in the barycentric frame precesses. This indicates a fixed orbital plane, but we also examine the focus and shape of the orbit.

\begin{figure}[p!]
\begin{center}
\includegraphics[width=7in]{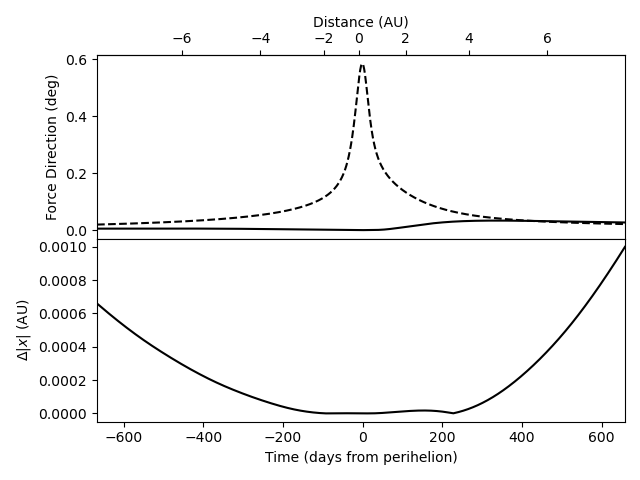}
\caption[]{
	\label{Fig6_FLdr_vsAU}
Change of force directions around a single perihelion encounter for shower \#1055. The change in directions of force (top), 
and cartesian distance between the computed orbit and a heliocentric elliptical orbit matching the numerical elements at perihelion (bottom) are shown versus distance from the Sun (top axis) and time (bottom axis). The top panel shows the angle between the direction of the net force and the rays from the comet to SSB (dashed) and the Sun (solid);  
the bottom panel shows the difference between a putative elliptical orbit about the Sun and the numerically computed orbit (solar motion is taken into account).  During the periapse crossing, the force points at the Sun, and the orbit is well approximated by an elliptical orbit with the Sun as focus.
}
\end{center}
\end{figure}

The top panel of figure~\ref{Fig6_FLdr_vsAU} shows the change in the direction of the force vector versus time (bottom axis) and heliocentric distance (top axis) in the days around perihelion, calculated from a perihelion crossing of shower \#1055.  The gravitational force from the mass of the Sun, eight major planets and Pluto was calculated at every time step of the high resolution simulations. The figure shows the angle between the force vector at each timestep with the vectors from the comet to the Sun (solid) and SSB (dot-dash). While inbound and well outside Jupiter's orbit, the comet sees a net force emanating from SSB. Inside Jupiter's orbit and near perihelion, the origin of the force swings about $0.5^\circ$ from SSB, and the force now is directed from within about $0.02^\circ$ of the Sun's center of mass. 

The bottom panel of Figure~\ref{Fig6_FLdr_vsAU} shows that an analytical elliptic orbit fit to the heliocentric elements at perihelion provides a good approximation to the numerical computations when the comet is near the Sun. The approximate elliptical orbit used for comparison was created using the orbital elements of the comet at perihelion that came out of the numerical computation.  To extend it away from perihelion, the orbital elements were fixed at their perihelion values, allowing only the true anomaly to change.  The figure shows the Euclidean distance between this approximate orbit and the numerically integrated orbit at times near perihelion. The fitted orbit approximates the numerical computations only within approximately 200 days of perihelion in this case.  As discussed below, the heliocentric orbit is not co-planar with either the incoming or exiting barycentric orbit.

\subsubsection{Location of the transition}

\begin{figure}[p!]
\begin{center}
\includegraphics[width=7in]{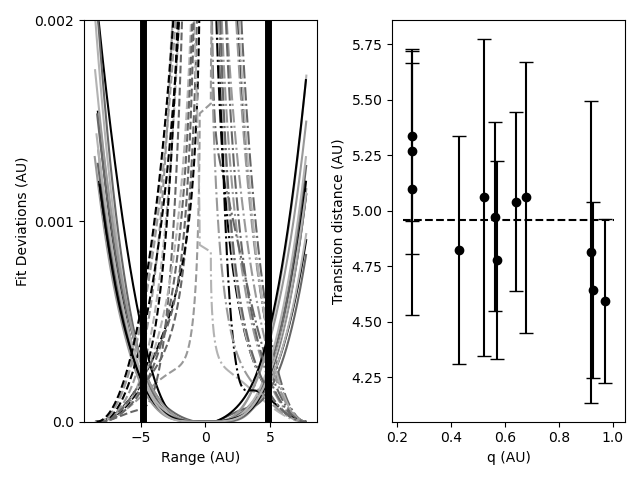}
\caption[]{
	\label{Fig7_5AU_Transition}
Graph showing the distance from the Sun where the transitions from barycentric to heliocentric orbits occur. The left-hand panel shows the Euclidean distance of the numerically integrated orbits for all perihelion crossings of shower \# 647 from theoretical orbits calculated from the elements at the preceding apehelion (dashed), perihelion (solid), and the succeeding apehelion (dot-dashed).  Thick vertical lines are placed at the average of the radii at which the distance to the heliocentric orbit becomes less than the preceding or succeeding barycentric orbit.
The right-hand panel shows the average values thus calculated for each shower in the study, along with the standard deviations.  The values are plotted versus $q$, indicating a possible movement inward from 5.2AU to 4.8AU as $q$ increases from 0.25 to 0.95.
}
\end{center}
\end{figure}

\begin{figure}[p!]
\begin{center}
\includegraphics[width=7in]{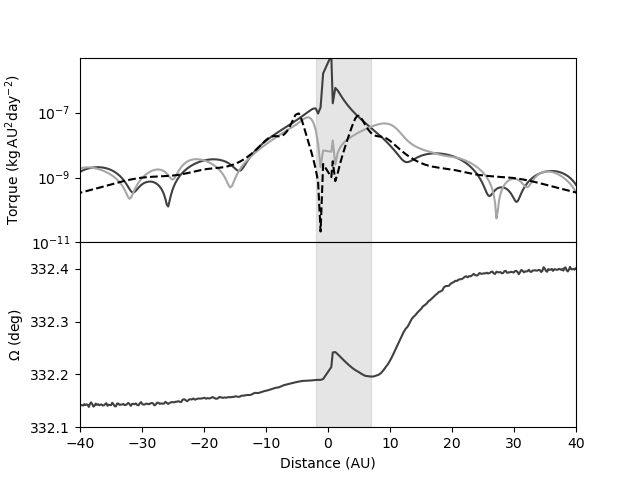}
\caption[]{
	\label{Fig10_TorqueOmega_vs_AU}
 Effects of planetary close encounters. The change in the longitude of ascending node, $\Omega,$ is shown versus distance (bottom), along with (top) the magnitudes torques due to the Sun (dark-gray) and planets (light-gray).  Also shown is the averaged torque predicted from spreading the mass of each planet around its orbit (dashed).  The region in which the comet is in orbit about the Sun is shaded. In this region, the oscillation of $\Omega$ can be seen.  $\Omega$ is seen to increase just before the comet enters the Sun's influence, due to planetary torques, and to again increase afterwards due to a close encounter with Jupiter.  The precession of $\Omega$ while under the Sun's influence can be seen in the hashed region.
}
\end{center}
\end{figure}

For all showers, the transition between barycentric and heliocentric orbits happens just inside Jupiter's orbital radius.
Figure~\ref{Fig7_5AU_Transition} shows a detail of the transition radius for shower \#839 across 14 perihelia (left), and the average transition radii across all twelve showers in our study (right).  In the left panel, we plot the Euclidean distance of the numerically integrated orbit through each perihelion from three paradigmatic orbits: (a) the barycentric orbit with elements at the preceding aphelion (dashed), (b) a heliocentric section of orbit computed from the elements at perihelion (solid), and (c) an outgoing barycentric orbit with elements corresponding to the succeeding aphelion (dot-dash).  As expected, the numerical orbits agree with the incoming barycentric orbit (a) until approximately 5AU, at which point they start to agree better with the heliocentric orbit (b).  Outbound, the heliocentric orbit stops being as good a representation at approximately 5AU as the barycentric orbit (c) becomes the best representation.  We compute the average radius at which the heliocentric orbit is closer to the integrated orbit both inbound and outbound.  The average values are shown as thick vertical lines on the left-hand panel.  We define this as the transition location, though we note that the orbit most closely fits an elliptical section about the Sun only deeper inside this region, when within 2 or 3AU from the Sun.

The right-hand panel of Figure~\ref{Fig7_5AU_Transition} shows the average transition region and standard deviation for the twelve showers we studied in detail.  These are plotted versus $q,$ which is the only orbital element they seem to show a relation with.  The relation is very slight, indicating that orbits with a lower perihelion distance may transfer to the Sun's control at slightly higher distances from the Sun than those with higher perihelion distances.


The transition between barycentric and heliocentric orbits takes place in the presence of a fairly constant torque from the planets, which causes the orbital precession.  We find that the planetary torque is generally delivered outside the region where the orbit is heliocentric, so that the constant precession may be somewhat separated from the perturbations to the orbital elements that are caused by the barycentric-heliocentric transitions.

Figure~\ref{Fig10_TorqueOmega_vs_AU} shows an example, from perihelion 7 of shower \#1191, of the net torque exerted by the planets and Sun versus time and distance (top), and the timewise evolution of the osculating angle of the ascending node, $\Omega,$ versus distance from perihelion (bottom).  Unlike the other elements, which change mostly when within the Sun's influence, $\Omega$ can be seen to change before the comet enters the Sun's sphere of influence.  The torques from the Sun (dark gray) and planets (light gray) are shown in the top panel, calculated from the ephemerides used in the numerical simulations.  The planetary torque was also estimated using the standard approximation of spreading each planet's mass evenly around its orbit (dashed).  The approximation agrees well with the actual numerical value when close to the Sun, although it diverges outside 15AU (not shown) because Gaussian averaging approach is not a good approximation for the large outer planets.  The initial and final parts of the change in $\Omega,$ at around 10AU, can be seen to line up with where the planetary torques maximize.  In this case, there is a post-perihelion encounter with Jupiter that produces the excess torque.  Far away from perihelion, the planetary torques are insignificant because the planetary forces are small; inside the orbit of Jupiter, torques fall off because the lever arms are becoming smaller.



\subsection{Element perturbations from a close encounter perspective}

\subsubsection{Changes in $a$ and $q$: gravitational boost and braking}

The most noticeable feature we see in sequences of orbits, visible even in Figure~\ref{Fig1_Orbital_Stream}, is the presence of stochastic jumps in period and semimajor axis.  The method of matched conics from planetary close-encounters is well studied, and applying it here we find that it explains the changes in period seen in our simulations as the result of gravitational boost and braking from the Sun, with respect to SSB.

There are many treatments of planetary close encounters.  \citet{Everhart1969} provides a simple, intuitive derivation of the perturbation to a comet's orbit during planetary close encounters, calculating the perturbations to the comet's momentum by accounting for the momentum imparted to the planet. For planetary close encounters, the comet enters a hyperbolic orbit about the planet for a brief period of time, while they both orbit the Sun.   We treat the orbital perturbation by the Sun in an analogous way, but with the comet falling temporarily into an elliptical orbit about the Sun, while they both orbit SSB.  In this case, the comet should experience an energy boost or braking which depends on the angle, $\alpha,$ between the Sun's velocity about SSB and the direction of the perihelion in the Sun's frame.  This is calculated as
$\cos\alpha =|\mathbf{\hat{v}}_\odot| \cdot \mathbf{\hat{e}},$
where $\mathbf{\hat{v}}_\odot$ is the unit vector in the Sun's velocity direction about the barycenter and $\mathbf{\hat{e}}$ is the eccentricity vector of the comet's orbit in the heliocentric frame.

\begin{figure}[p!]
\begin{center}
\includegraphics[width=7in]{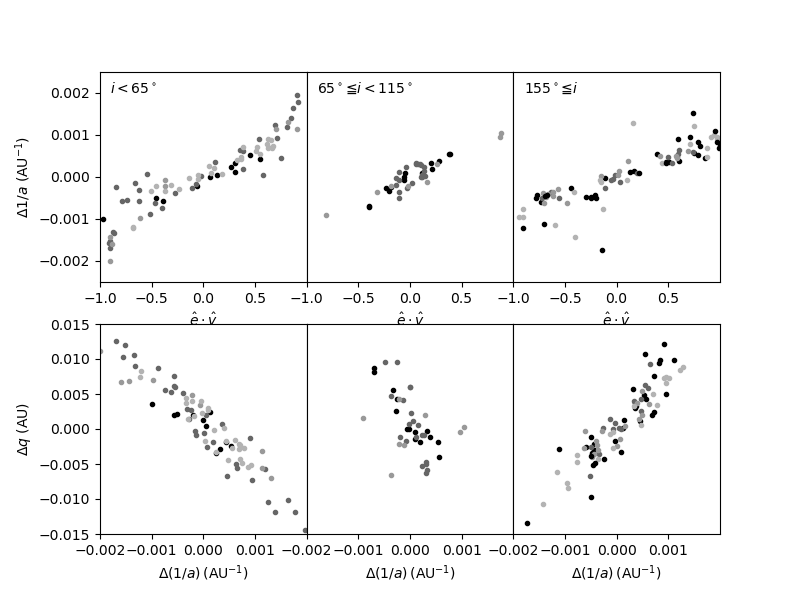}
\caption[]{
	\label{Fig8_dadq_vs_alpha}
Dependence of induced changes on the Sun state. Top: the change in kinetic energy, $\Delta(1/a)$, and change in perihelion distance, $\Delta q$ are plotted versus the cosine of the angle between the heliocentric eccentricity vector at perihelion and the Sun's velocity about SSB at perihelion, $\cos\alpha = {\mathbf p}\cdot{\mathbf v}$, as calculated across all perihelia for the twelve sample showers.  The plots are separated into prograde orbits (left), highly inclined orbits (middle), and retrograde orbits (left).  In all cases, the change in energy from one barycentric orbit to the next is strongly correlated with the Sun's motion during the encounter, in analogy with the usual concepts of gravitational braking or boost.  The change is less for highly inclined orbits, for which the directions of the Sun's motion about SSB and the comet's motion are more likely to be perpendicular.  Bottom: The change in perihelion distance also correlates strongly with $\alpha$, but with different signs for prograde and retrograde orbits due to the asymmetry induced by the orbit direction of the Sun.
}
\end{center}
\end{figure}

Figure~\ref{Fig8_dadq_vs_alpha} shows the change in energy and perihelion distance versus $\cos\alpha$ calculated across all perihelia from all twelve showers in the study.  The top panels display $\Delta (1/a),$ as a proxy for the orbital energy, as a function of $\cos\alpha$, with separate panels for prograde, highly inclined, and retrograde orbits. Within each panel, each dot represents the change in energy across one perihelion, and dots for each model comet are the same shade.  As expected for a close encounter, the energy change depends upon the angle $\alpha$ for the lowly-inclined orbits, both prograde and retrograde, though the correlation is slightly looser for lowly-inclined retrograde orbits.  When the comet catches up to the Sun, it loses energy;  when it approaches the Sun head-on, it gains energy.  For highly inclined orbits, we expect the effects to be small, because with the average orbital planes of the Sun and comet nearly perpendicular to each other, $\cos\alpha \ll 1.$ 

The bottom panels show the changes in the perihelion distance, $\Delta q,$ calculated across all perihelia of the twelve study comets.  The perihelion distance and total angular momentum, $\Delta |L|$, are directly related to the changes in the energy of the orbit (Table 2), and the directions of change are revealing as to the manner in which they also depend on the Sun's state during the close encounter at perihelion.
We show only the behavior of $\Delta q$ versus $\cos\alpha$, separated into groups by inclination.  $\Delta q$ and $\Delta|L|$ vary together, and both quantities vary differently with $\alpha$ for prograde and retrograde orbits.  For prograde orbits, $q$ and $|L|$ decrease when the energy decreases, but for retrograde orbits they both increase.  The reason for this is related to the asymmetry induced by the Sun's prograde motion about barycenter.  For prograde orbits, the Sun is systematically closer than SSB during the braking and farther when energy boost happens.  During braking there is a shortening of the lever arm and the angular momentum in the Sun's frame is less than the barycentric angular momentum, and during boost the heliocentric angular momentum is larger.  For retrograde orbits, the opposite happens.

For all orbits, if we consider the absolute value of the energy across perihelion, it increases (energy decreases) when the inbound comet is catching up to the Sun.  For prograde orbits, this occurs when the Sun is closer to the comet than the barycenter is, whereas for retrograde orbits this occurs when the Sun is farther from the comet than the barycenter is.  At the time when the Sun takes control, the velocity remains continuous from what it was during the barycentric portion of the orbit, but the new angular momentum is determined by the lever arm with the Sun.  For prograde orbits, $|L|$ decreases when the Sun is in the configuration when they lose energy ($\Delta(1/a)>0$), and increases when $\Delta(1/a)<0.$  Retrograde orbits behave in exactly the opposite manner.  The change in energy is always produced merely by the Sun's motion along the major axis of the orbit, and the fact that the Sun is prograde about SSB induces the asymmetry.

The perihelion distance, $q,$ changes with the magnitude of the angular momentum, because at perihelion $|L| = qv.$  For all orbits, the eccentricity decreases and semimajor axis becomes smaller with a decrease in energy, and the orbit becomes more circular.  But during this process the perihelion distance adjusts inwards for prograde orbits and outwards for retrograde.  Because the comet is expected to encounter the Sun with equally likely chances for braking or boost, the expectation value for the solar contribution to the change in net angular momentum or change in perihelion distance is zero.  These quantities may drift over time due to the Sun's impulses, but their random walk has an equal probability of forward and backward steps.

\subsubsection{Solar torque and perturbations to the angular elements}
  
Applying the close-encounter approach to the comet and Sun co-orbiting SSB requires some development not included in the usual planetary encounter treatment.  In that case the angular momentum of the planet about the Sun during the brief hyperbolic capture is negligible, whereas here the Sun carries significant angular momentum perpendicular to the ecliptic plane, and not aligned with the comet's angular momentum.  Torque from the Sun influences the angular orbital elements and contributes a dominant portion of the stochasticity seen in the nodal precession.

The precession of the barycentric angular momentum direction seen in Figure~\ref{Fig9_3D_Lvst} is a response to the out-of-plane torque delivered by the Sun.  The torque from the Sun determines the radius of the precession of the barycentric angular momentum vector.  If the Sun is above the orbital plane ($\mathbf{x}_{\odot}\cdot\mathbf{L} > 0$), the precession cycles through higher values of inclination, while if it is below the plane, the precession cycles through lower values but always in the same direction.  In a simplified treatment, the Sun delivers no torque to the comet during the time when the angular momentum about the Sun remains constant, rather the torque is delivered as the comet enters and leaves the heliocentric orbit.  The amount of torque applied is determined by the distance of the Sun relative to the comet's orbital plane at that time of the encounter. 

\begin{figure}[p!]
\begin{center}
\includegraphics[width=6in]{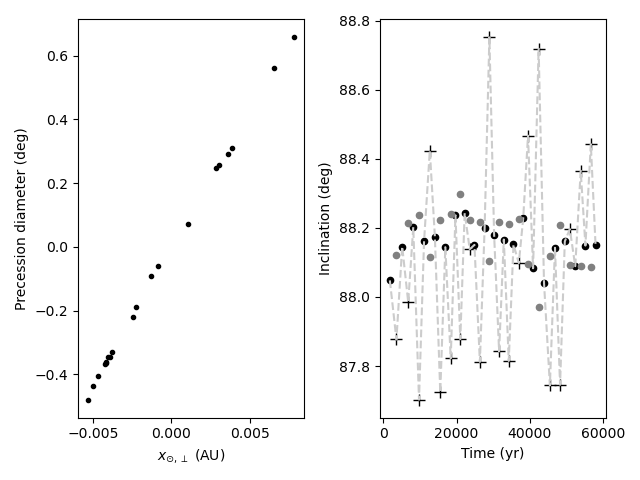}
\caption[]{
	\label{Fig11_DiL60kyr}
Magnitude of the angular momentum change versus Sun state. The left-hand panel shows the angular diameter of the circle traced out during the precession of the angular momentum is shown for all perihelia of shower \#1055, plotted versus the Sun's perpendicular distance from the orbital plane, $x_{\odot,\perp}$.  The value is positive for perihelia where $\Delta\Omega$ decreases (heliocentric inclination higher than barycentric), and negative when $\Delta\Omega$ is increasing. The magnitude of the precession, and the solar torque in barycentric coordinates, are directly proportional to the Sun's distance from the orbital plane.
The right-hand panel shows the inclination, $i$, versus times for apehelia and perihelia that occur during 60kyr of evolution for Shower \#1055.  Values in barycentric coordinates are connected by dashed lines, with the apehelion values marked in black crosses and perihelia in black dots.  The perihelion inclination in heliocentric coordinates is plotted as gray dots. These are seen to be in line with the barycentric values at aphelion (and hence during most of the orbit), whereas the value of the barycentric inclination at perihelion is caught halfway through its precession and is not representative of the orbital geometry during either portion of the orbit.
}
\end{center}
\end{figure}

The left-hand panel in Figure~\ref{Fig11_DiL60kyr} shows the dependence of the angular magnitude and orientation of the precession on the perpendicular distance of the Sun from the orbital plane for all perihelia calculated for shower \#1055.  The vertical axis shows the diameter of the precession (like that shown in Figure~\ref{Fig4_merge_dEdL}).  It can be approximately calculated as the perihelion inclination minus the inclination at the preceding aphelion in the SSB frame, so that it is negative when the perihelion inclination drops.  It is plotted here versus the distance of the Sun from the incoming barycentric orbital plane, $x_{\odot,\perp},$ which is calculated as the projection of the Sun's SSB position onto a unit vector in the direction of the incoming barycentric angular momentum and is negative when the Sun is 'below' the incoming barycentric plane. Capture into heliocentric orbit requires the orbital plane to change so that it includes the Sun.  The farther the Sun is from the incoming orbital plane at the time of transition, the larger the shift in orientation of the orbital plane and the larger the perturbation to the angular momentum direction.

The right-hand panel shows these transitions in context of 60kyr evolution of the shower.  The values of the inclination in barycentric coordinates are shown at each aphelion (black dots) and perihelion (black crosses), and are connected by a dashed line.  The aphelion values remain nearly constant, while the perihelion values, which are halfway through the precession, lie either above or below the line of aphelion values.  The ones which lie above are those for which $x_{\odot,\perp} > 0$.  The heliocentric inclination is shown at each perihelion (gray dots).  It lies in line with the drift of the apehelion barycentric angular momenta.  At perihelion, the Sun has taken over the orbit, and has done so in such a way as to keep the angular momentum direction somewhat constant.


The Sun's distance from the incoming barycentric orbital plane and its velocity with respect to the comet determine the perturbations to the angular elements.  Both $x_{\odot,\perp}$ and $v_\odot$ are characterized largely by its phase angle with respect to the nodal line, $\phi_{\odot}.$  In solar ecliptic coordinates, with our definitions for ${x}_{\odot,\perp}$ and $\phi_\odot$, to a good approximation ${x}_{\odot,\perp} \approx - |\mathbf{r}_\odot| \sin\phi_\odot\sin i.$  The elements $\Omega$ and $i$ change to accomodate the change in angular momentum, and $\omega$ appears to adjust in order to keep the eccentricity vector in a near constant direction while allowing apsidal precession.

\begin{figure}[p!]
\begin{center}
\includegraphics[width=7in]{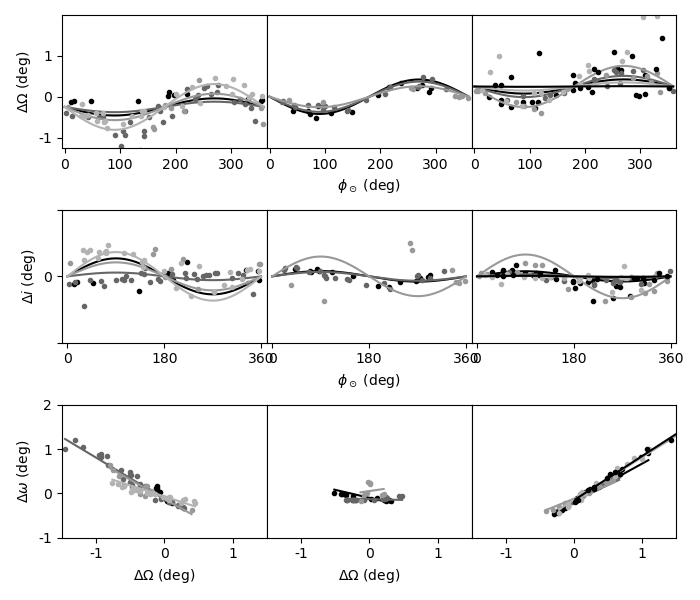}
\caption[]{
	\label{Fig12_dangelts}
Dependencies between the changes in the angular orbital elements. The top two rows show $\Delta\Omega$ and $\Delta i$ plotted versus $\phi_\odot$ for all perihelia of all comets in the study, separated into prograde (left), highly inclined (middle), and retrograde (right) orbits, and the bottom row shows $\Delta\omega$ plotted versus the change in nodal angle, $\Delta\Omega$.  Sinusoids are plotted atop the perturbations to $\Omega$ and $i$, whose amplitudes result from the standard deviation of the perturbations.  The sinusoid for $\Delta\Omega$ has an offset in mean due to the nodal progression, negative for prograde orbits and positive for retrograde.  $\Delta i$ is noisier than $\Delta\Omega.$  $\Delta\omega$ (bottom) changes opposite to $\Delta\Omega$, which reflects that the eccentricity vector does not change much during the perihelion transition.  The slope of $\Delta\omega$ versus $\Delta\Omega$ is $-\cos i$, as detailed in the text.
}
\end{center}
\end{figure}

Figure~\ref{Fig12_dangelts} shows the computed behavior of $\Delta\Omega$ and $\Delta i$ versus $\phi_\odot$ across all perihelia for our selected showers, and also the regression of $\Delta\omega$ versus $\Delta\Omega.$  As above, the cases are split into groups of prograde, highly inclined, and retrograde orbits. Each dot is the change evaluated across a perihelion crossing between two sequential barycentric orbits, with different shades for each shower. The top row of Figure~\ref{Fig12_dangelts} shows the variation of $\Delta\Omega$ superimposed with a sinusoidal fit, with amplitudes of $\sigma_{\Delta\Omega}$ derived below (Equation~\ref{eqn:sigma_elements1}) and an offset of $-0.15$ for positive $\cos i$ and $+0.15$ for negative $\cos i$. The amplitude is a quasi-analytical formula determined in the next section from Monte Carlo simulations varying the Sun's location.  The offset is approximate, and accounts for the nodal progression due to planetary torque.  

All of the prograde orbits show a net decrease in the nodal angle over time compared to the mean,  so that the expected value of the perturbations satisfies ${\rm E}(\Delta\Omega) < 0$.  The highly inclined ones show no net progression, while the retrograde orbits show an increasing nodal angle over time.  The amplitude of the effect of the perturbation due to the Sun can be larger than the nodal progression, but is superimposed atop it.

The middle row of Figure~\ref{Fig12_dangelts} shows the variation of $\Delta i$ with $\phi_\odot.$  The changes in $i$ are smaller and noiser than the changes in $\Omega.$ Nonetheless, a strong dependence on the Sun's angular position is observable.  We superimpose the sinusoids with amplitudes of $\sigma_{\Delta i}$ (Equation~\ref{eqn:sigma_elements2}, below).  The changes in inclination, $\Delta i$, are out of phase with the $\Delta\Omega$.  Both quantities respond alternately to the perturbations in the radius vector due to the solar position, $\mathbf{\hat r}_\odot$, as well as its perturbations in velocity due to $\mathbf{v}_\odot$.  Both the position and velocity of the Sun affect the direction of $\mathbf{L} = \mathbf{r}\times\mathbf{v}$.  The magnitude of the velocity perturbation out of the orbital plane is maximal near $\phi_\odot = 0$ or $180^\circ$, whereas the maximal out-of-plane perturbation due to $\mathbf{r}_\odot$ occurs near $\phi_\odot \sim 90$ or $270^\circ$.

The bottom row of Figure~\ref{Fig12_dangelts} shows the change in argument of perihelion between successive barycentric orbits, $\Delta\omega,$ versus $\Delta \Omega$.  Our numerical simulations and Monte Carlo simulations described below indicate that for any orbital configuration, the change in the argument of perihelion is nearly a linear combination of the change in line of nodes and change in inclination.  The linear coefficients depend in a complicated way on the orbital geometry, but singular value decomposition (SVD) applied to the triplet of values for all comets shows that to a good approximation there are only two degrees of freedom between these three variables.  

In general, $\Delta\omega$ varies with $\Delta\Omega$ nearly linearly with a slope approximately $(-\cos i)$.  It is important to note that the line containing $\Delta\Omega$ and $\Delta\omega$ does not pass through the origin but has an intercept.  Consequently, there is a range of values where, even though they move in lock-step with each other, $\Omega$ and $\omega$ are perturbed in opposite directions.  In our subsequent modeling, this allows them to random walk independently over time, even though their changes are tightly correlated at each perihelion.  

\begin{figure}[p!]
\begin{center}
\includegraphics[width=7in]{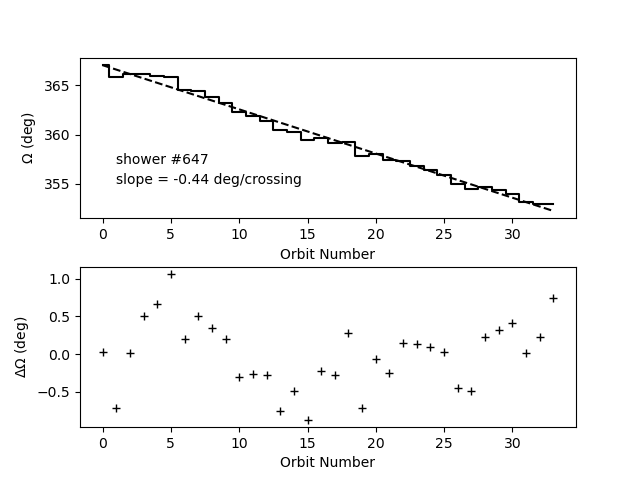}
\caption[]{
	\label{Fig13_Omegafit_wresid_nodalprecession}
Rotation of the nodal line from asymetry in induced kicks. The nodal regression of the prograde shower \#206 is shown versus orbit number, with a linear fit (top), and the residuals to the fit (bottom).  Although the planetary torques consistently pull the nodal line in one direction, the residuals show that the node steps in both directions.  The variance about the line is dominated by the Sun's state at perihelion.
}
\end{center}
\end{figure}

This also allows for a secular change in $\omega$ of about $0.1^\circ$ per orbit contrary to the direction of motion, presumably due to planetary torque.
Figure~\ref{Fig13_Omegafit_wresid_nodalprecession} shows the progression of the node for shower \#647 versus 33 perihelia crossings, spread over 60kyr.  The planets provide a systematic torque of $-0.44^\circ$ at each perihelion in this case, with residuals up to $1^\circ.$  The precession is a random walk that has a systematic component from the planets, a random contribution from the Sun, and a separate random component due to planetary close encounters.

\section{Statistics of perturbations resulting from Sun location}

The dispersion of meteoroid streams depends on the statistical behavior of perturbation to the orbital elements, $\Delta\Omega,$ $\Delta i,$ and $\Delta\omega,$ for an ensemble of similar orbits.  In this section, the dispersion growth will be  shown to depend on the standard deviations, $\sigma_{\Delta\Omega},\sigma_{\Delta i},$ and $\sigma_{\Delta\omega}$, of the perturbations across many solar encounters. We examine how the magnitudes of the perturbations at any crossing depend on the state of the Sun, and then how their statistical distributions depend on the statistics of the Sun's state as encountered by the comet.  

\subsection{Toy model}

We augment our numerical integration of orbits with a toy model of the physical interaction, and use an ephemeris-based Monte Carlo simulation of the Sun's state in order to validate the close-encounter model and quantify the expected distribution of perturbations to the angular elements.

The toy model expresses the assumption that the direction of the angular momentum vector instantly shifts when the comet crosses within 5AU of SSB, and that the comet then follows an elliptical orbit about this angular momentum axis until it is 5AU outbound, at which time it switches back to the barycentric frame. The physics embedded in this model is that momentum is transferred to and from the Sun during the entry and exit, and that there is no energy lost during the shift in angular momentum to be about the Sun. 

The state vector is converted from the incoming barycentric elements to the state vector, $(\mathbf{r},\mathbf{v})_1$, and thence to the outgoing barycentric state  vector, $(\mathbf{r},\mathbf{v})_2,$ and its associated elements.  Mathematically, following a similar derivation to that of \citet{Everhart1969}, the heliocentric section of orbit can be represented as a rotation expressed as 
\begin{eqnarray}
\mathbf{r}_2 & = & R(\mathbf{L}_\odot,\psi_r) (\mathbf{r}_1 - \mathbf{r}_\odot) + \mathbf{r}_\odot \\
\mathbf{v}_2 & = & R(\mathbf{L}_\odot,\psi_v) (\mathbf{v}_1 - \mathbf{v}_\odot) + \mathbf{v}_\odot,
\end{eqnarray}
where the $R(\mathbf{L}_\odot,\psi_{r,v})$ are the rotation matrices about the Sun's angular momentum vector of the angles appropriate for $\mathbf{r}$ and $\mathbf{v}$.  These angles follow from simple geometry as 
\begin{eqnarray}
\psi_r & = & 4\pi - 2\nu, \\
\psi_v & = & \pi - \zeta,
\end{eqnarray}
where $\nu$ is the true anomaly angle at $(\mathbf{r},\mathbf{v})_1$ and $\zeta$ is definied as the angle between the velocity vector and eccentricity vector, $\mathbf{e},$
such that $\zeta = \cos^{-1}(\mathbf{\hat{v}}_1\cdot\mathbf{\hat{e}})$.

This model doesn't include planetary torque or close encounters, and a further simplification in the model is that we assume that the Sun does not change state during this process.  Typically a comet takes several years to move from 5AU inbound to 5AU outbound.  This is up to 25\% of the Sun's co-orbital period with Jupiter and so it might move a substantial amount about SSB during this time.  Our results below, however, indicate that we can capture the main effects of the Sun close encounter seen in the numerical integrations to first order without accounting for that.

If we consider the Sun's orbit about SSB to lie in a disk of outer radius 0.010 AU (with even higher distances from SSB being rare), the inclination, $i$, nodal angle, and $\omega$ are sufficient to orient this disk with respect to the orbital plane.  The Sun's location within that disk is specified by $(r_\odot,\phi_\odot)$. Hence, the rotation is a function of $R = R(q, i, \omega; r_\odot,\phi_\odot).$  These angles implicitly reference the nodal vector and hence $\Omega,$ are invariant with respect to its value.

\subsection{Dependence on $q$}

\begin{figure}[p!]
\begin{center}
\includegraphics[width=7in]{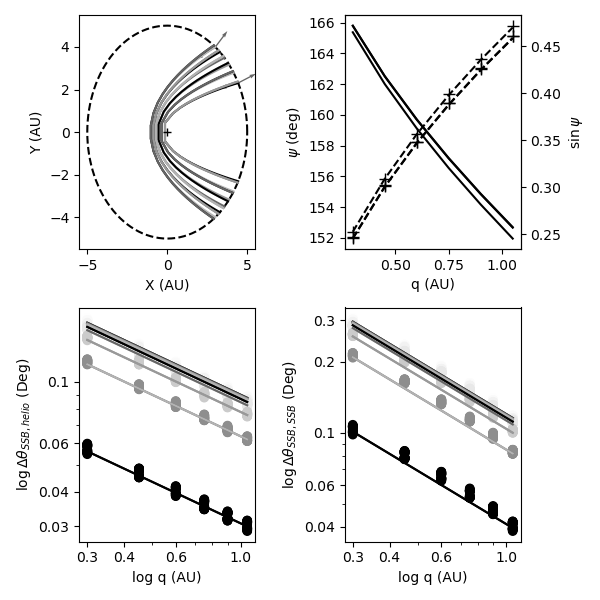}
\caption[]{
    \label{Fig14_qdependence}
Perihelion dependency of changes in the angular elements. The effect of $q$ on the orbital geometry is shown (top-left), and on the velocity-radius angle, $\psi$ at 5AU where the orbit becomes heliocentric (top-right).  Computed values from the toy model are shown versus $q$ in the bottom panels for $\Delta\theta_L$ (left) and the compound transition from one barycentric orbit to the next (right). 
In the top-left panel, the shape of orbits within 5AU is shown for 6 values of $0.3 < q < 1.05\,{\rm AU},$ and for semimajor axes with $a=50,500$ and 1000 AU.  For $q=0.3$ and 1.05, the velocity vector (light gray) and radius vector (black) of the incoming comet are shown. 
In the top-right, the angle $\psi,$ between $\mathbf{r}$ and $\mathbf{v}$ is plotted versus $q$ (left-hand axis, solid line), along with its sine (right-hand axis, dashed line) for the same grid of orbits.  Values of $\sin\psi = 0.59 q^{1/2}$ and $0.61 q^{1/2}$ are plotted in crosses.
The bottom panels are plotted in log scale to show that $\Delta\theta_L = \sin i\,q^{-1/2}$ for the single transition and $\Delta\theta_{L,tot} = \sin i\,q^{\sqrt{2}/2}$ for the compound inbound-outbound change.
}
\end{center}
\end{figure}

\citet{Jenniskens2023} and \citet{Pilorz2023} noted that the observed angular dispersions seem to grow as an inverse power of $q$.  The toy model reveals how this is caused by the geometry of the orbits, in that the angular change in the direction of the angular momentum during the transitions between barycentric and heliocentric coordinates and back, $\Delta\theta_L,$ grows as $q^{-\sqrt{2}/2}.$  These effects are illustrated with reference to  Figure~\ref{Fig14_qdependence}.

The top left-hand panel of Figure~\ref{Fig14_qdependence} shows the shapes of orbits within 5AU of the Sun in any orbital plane regardless of its orientation, for values of $q=0.30,0.45,0.60,0.75,0.90$ and $1.05$, and for $a = 50, 500, 1000$ AU.  For $q=0.30$ and $q=1.05$ the velocity vector is shown in light-gray and the position vector in black.  For orbits with semi-major axes $a > 50$ AU, the shape of the orbit within 5AU depends effectively only on $q$, with the near degeneracy in $a$ being what makes observational determinations of the semimajor axis difficult using only close-in observations.  It is clear that to a good approximation the true anomaly, $\nu$, at which the comet crosses the 5AU boundary depends entirely on $q$.  Likewise, $\zeta,$ the angle between the velocity and eccentricity vectors of the comet, as well as the angle between the radius and velocity vectors, $\psi \equiv \cos^{-1}(\mathbf{\hat{r}}\cdot\mathbf{\hat{v}}),$ both depend mostly on $q$ at the point where the Sun takes control. 

The top right-hand panel of 
Figure~\ref{Fig14_qdependence} shows how $\psi$ (solid, left axis) and 
$\sin\psi$ (dashed, right axis) vary with $q$ just inside Jupiter's orbit, and that this variation does not depend on the value of the semi-major axis.  Curves are shown for $a=50,500,1000$ AU, but the $500$ and $1000$ AU curves lie nearly atop each other.  All long period orbits trend to this value.  The curves begin to diverge from this asymptotic behavior for shorter orbits.  For $a=50$ the value of $\psi$ has changed by less than a percent, while the trend remains the same.  For all orbits, $120^\circ < \psi < 160^\circ,$ and in this region, the $\sin\psi \propto q^{1/2},$ which is plotted in crosses.  This appears to be the main source of the $q$ dependence of the perturbations.

The change of orbit focus from SSB to the Sun induces a perturbation to the angular momentum direction due to the fact that neither the radial vector nor velocity vector for the Sun is generally in the orbital plane.  Because of a lever-arm type effect in keeping the angular momentum perpendicular to both velocity and radius vector, the perturbations to $i$ and $\Omega$ are generally larger than the angular perturbations to $\mathbf{r}$ and $\mathbf{v}.$  For simplicity, we first analyze the effect of perturbations to $\mathbf{\hat{r}},$ while pretending that $\mathbf{v}$ remains constant.  The Sun's location away from SSB requires small shifts in the azimuth and declination of $\mathbf{\hat{r}}$.  The azimuthal shift does not affect the angular momentum direction, but the declination component does.  This component causes a small vertical angular perturbation to the radius vector on the order of $\theta_r = (r_\odot/|r|)\sin i \sim 0.002\sin i$ radians.  

Without loss of generality, we adopt a coordinate system in the orbital plane with $\mathbf{\hat{v}}\equiv(1,0,0),$ $\mathbf{\hat{L}}\equiv(0,0,1),$ and $\mathbf{\hat{r}}\equiv(\cos\psi,\sin\psi,0)$.  The angular momentum must be perpendicular to both $\mathbf{\hat{v}}$ and $\mathbf{\hat{r}}$, so that in this coordinate frame we can write the perturbed angular momentum as $\mathbf{\hat{L}}^\prime = (0,\sin\theta_L,\cos\theta_L),$  where $\theta_L$ is the change in declination of $\mathbf{\hat{L}}.$  We evaluate the value $\theta_L$ must have in order for the angular momentum vector to also be perpendicular to the perturbed radial vector by requiring that $\mathbf{L}^\prime\cdot(\mathbf{r+\delta r}) = 0.$ 

We assume first that $\mathbf{\hat{r}}$ has rotated by $\delta\theta_r$ in the plane containing $\mathbf{\hat{L}}$ and $\mathbf{\hat{r}}$.  The rotation axis is the vector $\mathbf{a} \equiv (\cos(\psi-90),\sin(\psi-90,0) = (\sin\psi,-\cos\psi,0).$ The rotation matrix of the angle $\theta_r$ about $\mathbf{a}$ is
\begin{equation}
\mathbf{R} = \left(
\begin{matrix}
\sin^2\psi + \cos^2\psi\cos\theta_r & -\sin\psi\cos\psi(1-\cos\theta_r)         & \cos\psi\sin\theta_r \\
-\sin\psi\cos\psi(1-\cos\theta_r)   & \cos^2\psi + \sin^2\psi\cos\theta_r    & \sin\psi\cos\theta_r\\
-\cos\psi\sin\theta_r & -\sin\psi\sin\theta_r & \cos\theta_r
\end{matrix}\right).
\end{equation}
The result of applying this matrix to $\mathbf{\hat{r}}$ is
\begin{equation}
\mathbf{\hat{r}}^\prime = \mathbf{R}\mathbf{\hat{r}} =
\left(\begin{matrix}
\cos\psi\cos\theta_r\\
\sin\psi\cos\theta_r\\
-\sin\theta_r
\end{matrix}
\right)
\end{equation}

Requiring $\mathbf{\hat{r}}^\prime$ to be perpendicular to $\mathbf{\hat{L}}^\prime$ gives us the constraint
\begin{equation}
(0,\sin\theta_L,\cos\theta_L)\cdot(\cos\psi,\sin\psi\cos\theta_r,-\sin\theta_r)  =  0,
\end{equation}
reducing to
\begin{equation}
\tan\theta_L  =  \frac{\tan\theta_r}{\sin\psi}.
\end{equation}
Using small angle approximations and our known behavior of $\sin\psi$ from Figure~\ref{Fig14_qdependence}, the maximum $\theta_L$ for any inclination orbit is then
\begin{equation}
\theta_L = \frac{0.002 \sin i}{\sin\psi} \propto \frac{\sin i}{q^{1/2}}.
\end{equation}

For an accurate treatment, one would have to account for the fact that the Sun's velocity also perturbs $\mathbf{v}$, and that the angle $\psi$ changes slightly due to the azimuthal perturbations not considered here.  These effects should not change the principal conclusion.  If we apply the above argument to the change of angular momentum when exiting the Sun's sphere as well as when entering it, then the total change in angular momentum will be the composition of the two changes.  These changes both have a $q^{-1/2}$ dependency, but they are in different directions so that the net change in direction is a vector sum.  The change in angular momentum direction appears to be a random variable with a standard deviation of $q^{-\sqrt{2}/2} = q^{-.707}.$

The bottom panels of Figure~\ref{Fig14_qdependence} show the dependence of the standard deviations in angular momentum change versus $q$ for the single transition from incoming barycentric to heliocentric frames (left) and the compound transition from the inbound to outbound barycentric frame (right), which involves transfer to and from the heliocentric orbital plane.  The calculations were performed for a grid of orbits with $0.30 < q < 1.05\,{\rm AU},$ $20^\circ < i < 170^\circ,$ and $\omega$ ranging from $0^\circ$ to $360^\circ$.   For each geometry, the angles for $\theta_{L}$ and the compound change, $\theta_{L, \rm{tot}}$, were calculated as resulting from 5000 random Sun locations, and the standard deviations tabulated and shown here.  The values are plotted versus $q$ and shaded as $\sin i$.  In the left-hand panel, the function $\Delta\theta_L = 0.1 q^{-0.5}\sin i$ is plotted for each $(i,\Omega)$ pair.  In the right-hand panel, the function $\Delta\theta_{L,\rm{tot}} = 0.1 q^{-.707}\sin i$ is shown.  The perturbation, $\theta_L$, is the hypotenuse of a spherical right triangle whose sides are $\Delta i$ and $\Delta\Omega$.  The simulations validate that while the change in angular momentum direction associated with the transitions in or out of the heliocentric frame each go as $q^{-1/2},$ their combined changes track as $q^{-\sqrt{2}/2}.$

\subsection{Statistics of solar induced angular element perturbations}

We use the model above with Monte Carlo realizations of the Sun's state in order to predict the standard deviations of the perturbations to the angular elements.  For each orbit in a grid of orbital geometries corresponding to our 12 study showers, we simulate 5000 Sun states sampled at random from the ephemeris during a 10kyr span of time around a typical perihelion crossing.  We take these as examples of the probability distribution of the Sun's state.  We use the toy model to map the inbound orbital elements onto the outbound orbital elements at 5AU for each orbital geometry to get 5000 realizations of $\Delta\Omega, \Delta i,$ and $\Delta\omega,$ from which we estimate $\sigma_{\Delta\Omega},\sigma_{\Delta i},$ and $\sigma_{\Delta\omega}$ using the standard unbiased maximum likelihood estimator.

\begin{figure}[p!]
\begin{center}
\includegraphics[width=7in]{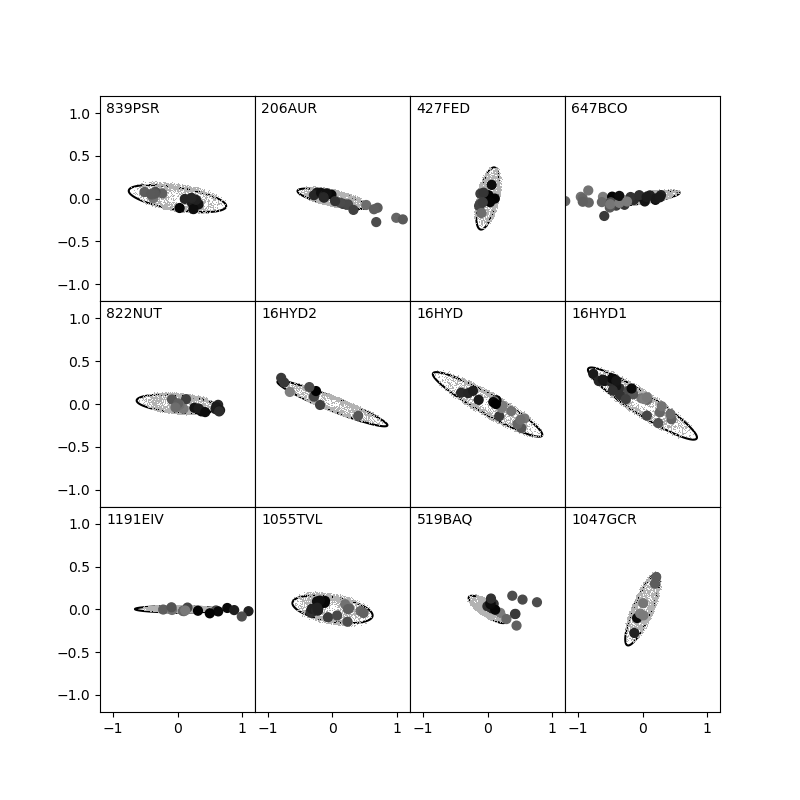}
\caption[]{
  \label{Fig15_dinclination_v_dOmega}
Dependency of changes in $i$ and $\Omega$ with the Sun's state. The theoretical and numerically computed co-variation of $\Delta i$ and $\Delta\Omega$ are shown as a function of Sun location for all comets in the study.  Each panel shows $\Delta i$ versus $\Delta\Omega$ for a study comet, and the axes on all panels are in degrees, with the horizontal axes being $\Delta\Omega$ and the vertical axes being $\Delta i$. Theoretical values computed using the solar close-encounter model are shown in black dots and as black curves.  The curves represent the $(\Delta i,\Delta\Omega)$ that would occur if the Sun were on a circular orbit about SSB of radius 0.01 AU.  The small dots show values that would occur using 1000 randomly sampled locations of the Sun taken from the numerical integrations for each shower, during the half-period before a selected perihelion.  The large dots are the values actually computed, shaded by $\phi_\odot$.
}
\end{center}
\end{figure}

Figure~\ref{Fig15_dinclination_v_dOmega} shows the behavior of the toy model together with the perturbations calculated from our numerical integrations, applied to the perturbations to the direction of the angular momentum vector.  The toy model appears to capture much of the behavior displayed in the integrations.
Each panel shows the perturbations to the angular momentum direction, $(\Delta\Omega,\Delta i)$, for an ensemble of random Sun locations applied to a randomly selected perihelion of each of the study showers with the computed perturbations from the numerical integrations of that shower.  The small gray dots shown in each panel of Figure~\ref{Fig15_dinclination_v_dOmega} are the $(\Delta\Omega,\Delta i)$ values predicted for 5000 randomly sampled locations of the Sun. The black circular curves represent the theoretical loci of $\Delta i$ and $\Delta\Omega$ that would result were the Sun on a circular orbit of radius $0.01$ AU in the solar ecliptic plane.  The shaded larger dots are the $\Delta i$ versus $\Delta\Omega$ computed from the numerical integrations, with each dot corresponding to a particular perihelion.  They are shaded by the solar angle with respect to the nodal line, $\phi_\odot.$  They are all inside the solid ellipses, because the Sun rarely strays as far as $0.010$ AU from SSB.  

The Monte Carlo simulations do not include the torque from the planets that leads to nodal progression, and close observation of Figure~\ref{Fig15_dinclination_v_dOmega} reveals that the computed results from the prograde showers are all slightly offset to the left from the toy model ellipses, while those from the retrograde offset to the right.  This is the same as the offset in sinusoids seen in Figure~\ref{Fig12_dangelts}.  The relative size of the systematic offset to the other variations offers a graphical assessment of the relative size of the planetary perturbations causing the nodal progression to the solar perturbation.

The Monte Carlo simulations also do not include planetary close encounters. The perihelia most affected by planetary close encounters are seen in the figure to be those having shaded dots that lie outside the ellipses defining the theoretical loci.  The fact that most of the perihelia have $\Delta i$ and $\Delta\Omega$ in the ranges predicted by the simple model indicates that within our sample, planetary encounters close enough to influence the dynamics are relatively rare.  Showers \#206 and \#647, both at relatively low inclinations, both have a number of successive orbits where the comet repeatedly encounters Jupiter, either outbound or inbound, creating a larger $\Delta\Omega.$ 

The size and orientation of the ellipses in Figure~\ref{Fig15_dinclination_v_dOmega} depend upon the orbital geometry.  Estimates for the expected ranges of values of $\Delta\Omega$ and $\Delta i$ versus orbital geometry may be inferred by regressing the ellipse orientation and major and minor axis lengths versus $i, \omega$ and $\Omega$, and using the horizontal and vertical extent of the ellipse as a proxy for the range of kick sizes to be expected.  The expressions for the standard deviations estimated by fitting the ellipse behavior as a function of orbit geometry are the same as those produced by directly computing the standard deviations of the simulated perturbations produced with varying orbital geometries.

The probability distribution of the Sun's radial distance from SSB is a wide mono-modal function that has a level top between 0.03 and 0.06 AU and falls off steeply both inside and outside that.  The distribution of its azimuthal location is uniform over long periods of time.  The isocontours of this distribution would appear in Figure~\ref{Fig15_dinclination_v_dOmega} as a series of concentric ellipses in the  $(\Delta\Omega,\Delta i)$  plane, with maxima near 0.05AU that would map to an ellipse about halfway along the major and minor axes. The values of $(\Delta\Omega,\Delta i)$ from the numerical distributions and from the perihelia integrated numerically are seen in Figure~\ref{Fig15_dinclination_v_dOmega} to lie mostly in this region.

\begin{figure}[p!]
\begin{center}
\includegraphics[width=7in]{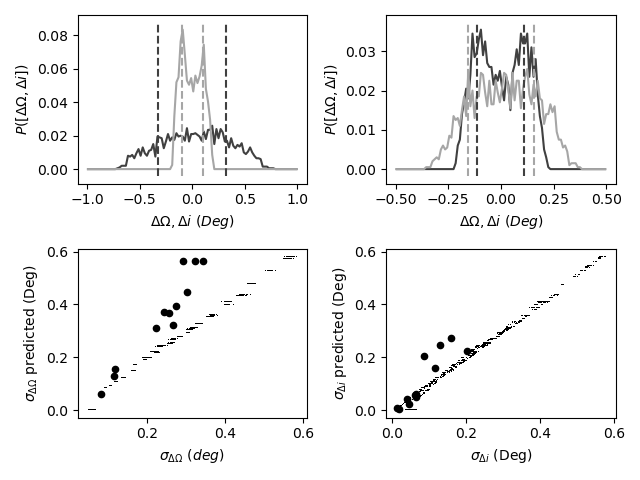}
\caption[]{
  \label{Fig16_SigmaOmegai_Regressions}
Sun Close-Encounter model of induced changes in $i$ and $\Omega$. The top panels show the marginal distributions of $\Delta\Omega$ (dark-gray) and $\Delta i$ (light gray) found from 2-dimensional distributions shown in Figure~\ref{Fig15_dinclination_v_dOmega}.  The left upper panel is calculated from shower \#839, which has a nearly horizontal ellipse with $\Delta\Omega$ along the major axis, and the right panel from shower \#427 which has a more vertical ellipse with $\Delta i$ across the major axis.  In each case, the variable aligned with the major axis shows a monomodal wide distribution, and the variable whose marginal integrates along the major axis shows a bi-modal distribution.  This is because the probability contours within the ellipse are concentric ellipses representing deformations of circles of constant solar radii.  The probable Sun location peaks around 0.05AU so that the net distribution has a maximal probability about halfway along the major and minor axes.  The bottom panels show how the standard deviations of the kicks, $\sigma_{\Delta\Omega}$ (left) and $\sigma_{\Delta i}$ (right), regress against the orbital parameters, $q,i$ and $\omega$.
}
\end{center}
\end{figure}

Figure~\ref{Fig16_SigmaOmegai_Regressions} shows the marginal distributions of $\Delta\Omega$ and $\Delta i$ for showers \#839 and \#427 (top), along with the general dependence of the standard deviations on the orbital parameters (bottom).
On the top, the marginal distribution of $\Omega$ is shown in black, and that of $i$ is shown in gray. Showers \#839 and \#427 were chosen for display because they have ellipses whose major axis is oriented respectively parallel to $\Delta\Omega$ and $\Delta i$ (see Figure~\ref{Fig15_dinclination_v_dOmega}). 
The marginal distribution when looking perpendicular to the major axis the ellipse has a single wide peak, where at each value of $\Delta\Omega$ (or $\Delta i$), one is seeing 'across' the ring of highly probable points.  The marginal distribution viewed along the major axis is thinner.  It looks 'along' the ring of high density points, and appears to have a twin peaked distribution.

The actual functions for $\sigma_{\Delta\Omega}(i,\omega,q)$ and $\sigma_{\Delta i}(i,\omega,q)$ are complicated.  Fitting the toy model Monte Carlo simulations  across our grid of orbital elements in $q,i$ and $\omega$ results in an orbital-element dependence of the perturbations which can be expressed as
\begin{eqnarray}
\sigma_{\Delta\Omega} &\approx & 0.25|\sin\omega|q^{-0.707} \label{eqn:sigma_elements1}\\
\sigma_{\Delta i}     &\approx & 0.25 |\cos\omega| (sin \: i) q^{-0.707}\label{eqn:sigma_elements2}\\
\sigma_{\Delta\omega} & \approx &   |\cos i| \sigma_{\Delta\Omega},\label{eqn:sigma_elements3}
\end{eqnarray}
but the fit is very loose.

The bottom panel of Figure~\ref{Fig16_SigmaOmegai_Regressions} shows the standard deviations predicted from the simplified formulae in Equations~\ref{eqn:sigma_elements1}-~\ref{eqn:sigma_elements3} versus the standard deviations actually produced in simulations of the toy model, as small black dots.  The agreement indicates merely how tight the correlations in the equations are.  The large dots in Figure~\ref{Fig16_SigmaOmegai_Regressions} are the predicted versus computed standard deviations from each of our 12 sample showers. The computed standard deviations are calculated from a small number of perihelia crossings for which planetary encounters were not important, but in spite of the small number statistics it is apparent that the variances from the simulations are less than those derived from the toy model.  

Examination of Figure~\ref{Fig15_dinclination_v_dOmega} indicates that the standard deviations of the computed simulations are indeed smaller than those of the toy model.  The dispersions seen in the numerical integrations might be expected to be smaller than those obtained from a full sampling of all possible Sun states, because even during a typical 60kyr evolution of a comet orbit not all Sun states are sampled.  Changing the numerical constant from $0.25$ to $0.15$ in Equations~\ref{eqn:sigma_elements1} and~\ref{eqn:sigma_elements2} makes a better fit to what we evaluate from numerical integrations.

\section{Discussion}


The results presented above demonstrate that an orbiting comet executes a sequence of barycentric orbits, and that the changes in orbital elements from one orbit to the next occur near perihelion and depend to a large degree on the state
of the Sun relative to SSB as it is encountered. A toy model constraining only angular momentum shows good overall agreement with the perturbations to the angular orbital elements seen in detailed numerical integrations, although it
predicts variances roughly a factor of two higher than the numerical integrations indicate actually occurs. In spite of this discrepancy, we assume that the physical picture presented above is substantially correct in broad strokes.

Using the above model, and assuming that the state of the Sun encountered by an LPC on each approach is random, we can attempt to predict the dispersion growth of a stream over time using a simple random walk model for the evolution of orbital elements during the long period time evolution of the comet. We build this model below, show how to extend it to the associated meteoroid stream, and show that its behavior applied to the meteoroids resulting from a single perihelion crossing agrees with properties of the dispersion
growth that we have seen in previous long term numerical simulations. Before doing that, we discuss the evolution of streams in order to put the problem in context.

\subsection{Growth of dispersion for meteroid streams and LPC shower age}

Meteor observations most precisely measure the angular orbital elements: $i, \omega,$ and $\Omega,$ whereas the semi-major axis and eccentricity are not precisely enough measured to see the effect of age. The perihelion distance, $q$, is measured accurately as well, and while it is generally stationary in value, the dispersion in $q$ can also be used as a measure of stream age. 

\citet{Jenniskens2023} has regressed an approximate age for showers from the observed dispersions in inclination, argument of perihelion and node, after taking into account the observational error in measuring the inclination and argument of perihelion. The dispersion of the node is precisely measured as the node relates closely to the time of the meteor. To minimize observational uncertainty in any of these angles, the calculated dispersions were taken together as:
\begin{equation}
    \sigma_{tot} = \left(\sigma_\omega^2 + \sigma_i^2 + [\sigma_\Omega  \sin(\delta_v)]^2 \right)^{1/2},
    \label{eqn:sigma_rss}
\end{equation}
where $\delta_v$ is the the attack angle of Earth's path through the stream that affects the observed dispersion in node. The attack angle is expressed as 
\begin{equation}
 \delta_v = 180^\circ - \cos^{-1}\left(\sin(\lambda + \lambda_\odot) \cos(\beta)\right),
    \label{eqn:delta_v}
\end{equation}
where $\lambda , \beta$ are the ecliptic longitude and latitude of the radiant, respectively, and $\lambda_\odot$ is the solar longitude. For each shower studied here, these ecliptic radiant coordinates are reported in \citet{Jenniskens2023}. 

It was found that to first order, for ages from about 10,000 to 120,000 years,
\begin{equation}
    {\rm Age}(yr) = 12,000 \: q \: \sigma_{tot},
    \label{eqn:sigma_age}
\end{equation}
with the implication that $\sigma_{tot}(t_{yr}) = t_{yr}/(12000q)$ with $t_{yr}$ measured in years and $q$ measured in AU. That these age estimates were perhaps 27$\%$ higher than results from other models was a factor that prompted this investigation.
Although the bulk of LPCs (250 - 4000 year orbits) have a dispersion growth that fits well by Equation~\ref{eqn:sigma_age}, shorter period (10-160 y) Mellish type showers required a higher slope of $t/(6000q)$.  

Long-time numerical simulations run to validate these studies did indicate an inverse $q$-dependence of the growth, but also found that if $a$ and $e$ are adjusted to vary the period of the comet in numerical simulations while leaving other orbital elements the same, then the slope of $\sigma_{\rm tot}(t)$ varied as well \citep{Pilorz2023}.  The simulations also indicated that when tabulated in terms of orbit crossings, $n$, instead of time, the dispersion grows at the same rate for long and short period comets, both roughly as $n^{1/2}$.   

\begin{figure}[p!]
\begin{center}
\includegraphics[width=7in]{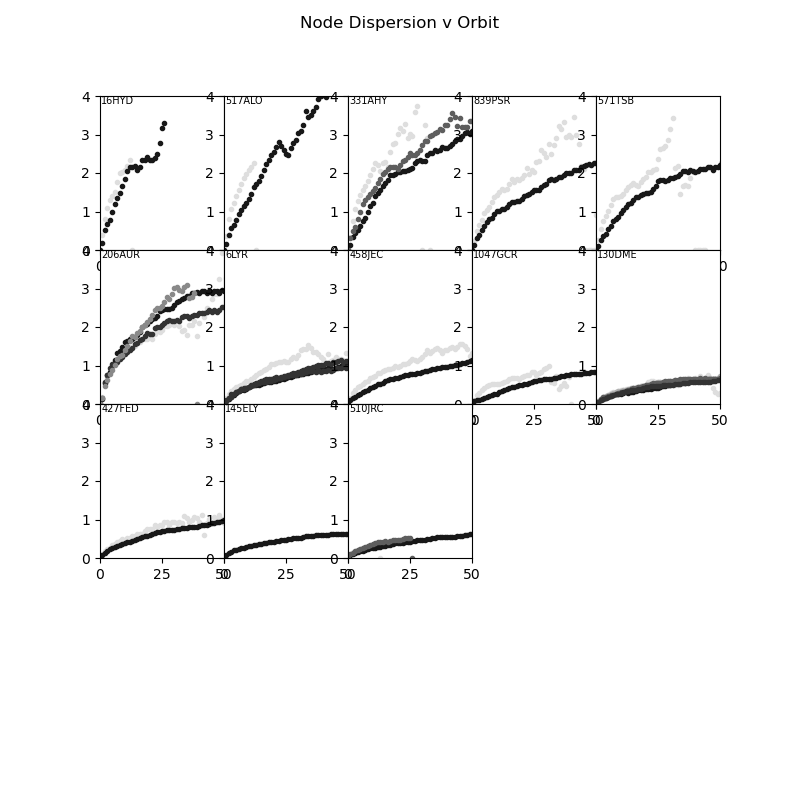}
\caption[]{
    \label{Fig19_From_ACM}
Growth of dispersion in $\Omega$ as a function of the number of completed orbits. The showers are arranged in order of increasing $q$ from top-left to bottom right, where $ 0.3 \lessapprox q \lessapprox 1$. The dispersion in nodal angle, $\Omega,$ versus number of orbits has been calculated from long term 60kyr+ numerical simulations based on 11 showers within and outside of the showers used in this study.  For each shower, at least two different orbital periods, usually 250 and 4000 years, were simulated by adjusting $e.$  The dispersion growths for these disparate periods lie nearly atop each other when plotted versus orbit, although there is substantial variation due to the stochastic nature of the orbits.  The growth of dispersion is faster for orbits with small perihelion distances, and approximately follows $\sigma_\Omega(n) \propto q^{-0.7} n^{1/2},$
where $n$ is the number of orbits completed.
}
\end{center}
\end{figure}

That can be seen in Figure~\ref{Fig19_From_ACM}, which shows the results of numerical simulations of the dispersion of long term evolution of meteoroid streams for a number of different showers.  Each simulation follows 1000 particles that were ejected during a single perihelion crossing. Two or three variants of each orbit were run in which the eccentricity was adjusted to produce periods of 250 (near the lower limit for an LPC) and 4000 years (upper limit at which the stream would be dense enough to be detected), and sometimes also 1000 years, and all of these were followed for 60-100kyr of evolution. The dispersion in the nodal angle, $\sigma_\Omega,$ is shown here versus orbit number for all of the first fifty orbits.  

In Figure~\ref{Fig19_From_ACM}, the showers are arranged in the order of their initial value of $q$ from lowest to highest, going from top left to bottom right.  For each shower, the curves for 250 year (dark-gray) and 4000 year (light-gray) orbital periods lie largely atop each other, consistent with the growth depending primarily on the number of encounters the particle has experienced with the inner solar system.  When plotted versus time, the shorter period simulations display a steeper slope than do the longer period ones, but versus orbit number the two curves lie fairly atop one another.  The growth with orbit number, $n$, is not linear but appears to increase roughly as $n^{1/2}.$  There is a clear trend in that showers with higher perihelion distance disperse more slowly.

The primary issues this work set out to address were to understand the reason for the dependence of dispersion growth on the number of orbits along with its $n^{1/2}$ behavior, the reasons for the inverse $q$ dependence, and the physical drivers of the dispersion.  We have discussed the $q$ dependence above, and now have the tools to address the dispersion growth.

\subsection{Meteoroid cloud dispersion as a random walk}

Physically, the comet receives an impulse at each perihelion that determines its outbound orbit.  The orbital period changes due to gravitational boost or braking, and the orientation changes due to the slight difference in the direction of the angular momentum vector about the Sun from that about SSB. The comet then follows the perturbed orbit until its next perihelion crossing, where it receives another impulse. Mathematically, the secular evolution of the orbit proceeds as a random walk in the phase space of the orbital elements, with each perihelion crossing contributing a random kick to the elements. The chance location of the Sun is the source of randomness. The perturbation to $\Delta (1/a)$ determines the step's duration, and the change in other elements determines the step's direction.  The walk is Markov, because each step depends only upon the incoming elements.  Planetary torques are present and deliver a systematic forcing that results in nodal and apsidal precession.  Planetary encounters introduce another stochastic component which we will not consider here.

To formalize this process, define the element vector as $\mathbf{S}(n) \equiv (a, q, i, \omega, \Omega)$ as the set of orbital elements after the $n$th perihelion crossing. Each element is $\mathbf{S}_{n,j}$, where j delineates the element, and is between 0 and 4 inclusive.  Then for any orbit, $k$, and element, $j$,
\begin{equation}
  \mathbf{S}_{k,j} = \mathbf{S}_{k-1,j} + \Delta\mathbf{S}_{k,j} 
                   \approx \mathbf{S}_{k-1,j} + (\Delta_\odot S + \Delta_{\rm pl}S + \Delta_\tau S)_{k,j}
  \label{eqn:markovstate}
\end{equation}
The $\Delta\mathbf{S}_{k,j}$ are the kicks received by each orbital element as the comet passes through perihelion, including the solar perturbations ($\Delta_\odot S$), planetary close encounters ($\Delta_{\rm pl} S$), and systematic solar system torques,$\Delta_\tau S$  At each step, the perturbations to the various elements can be thought of as random variables that depend on the probability distribution, $P(r_\odot,\phi_\odot),$ of the Sun's location.  

Neglecting the planetary close encounters, we formulate the planetary torque as being constant with no variation, and the perturbations due to the Sun as varying symmetrically about zero.  The expectation and variance of the forcing terms are then
\begin{eqnarray}
    {\rm E}[\Delta_\odot S] & = & 0 \\
    {\rm V}[\Delta_\odot S] & = & \sigma^2_{\Delta S}\\     
    {\rm E}[\Delta_\tau S] & = & \Delta_\tau S\\
    {\rm V}[\Delta_\tau S] & = & 0.    
\end{eqnarray}
The solar forcing has an expected value of zero at any encounter, but a variance determined by the variance of the forcing terms;  the torques are systematic and have no variance, but do have an expected value that produces orbital progression.

We can use the statistics of the forcing to evaluate the expectation and variance of the orbital elements as a function of the number of orbits completed.  If we consider the ensemble after each member has completed $n$ orbits, the statistics of the orbital elements are:
\begin{eqnarray}
{\rm E}\left[\mathbf{S}_{j}(n)\right]& = & \mathbf{S}_{0,j} + {\rm E}\left(\sum_{i=1}^n \Delta\mathbf{S}_{i,j} \right)
                 = \mathbf{S}_{0,j} + n\Delta_\tau S_j 
                 \label{eqn:expec_over_orbit}\\
{\rm V}\left[\mathbf{S}_{j}(n)\right] & = &
    \sum_1^n \sigma^2(\Delta_\odot\mathbf{S}_j) 
     =  n \sigma^2_{\Delta j}
    \label{eqn:var_over_orbit}
\end{eqnarray}
The expectation value of the state vector after $n$ perihelia is merely $n$ times the systematic planetary torque it receives at each perihelion.
The variance describing the uncertainty in element grows linearly with $n$, and is the sum of the variances of the perturbations to the angular elements at each perihelion, $\sigma^2_{\Delta\Omega},$ $\sigma^2_{\Delta i}$, and $\sigma^2_{\Delta\omega}.$  Taking the square root of Equation~\ref{eqn:var_over_orbit}, the dispersions grow generally as the square root of the number of orbits completed, $\sigma_{\mathbf{S}_j}(n) \propto n^{1/2}$.  This is true of the root-sum-square as well as of the perturbations of the individual elements.

We derived Equation~\ref{eqn:var_over_orbit} to express the uncertainty in the state of a comet after $n$ orbits, but a simple line of reasoning allows us to apply it to the dispersion of the meteor stream.    Over a few hundred thousand years we expect that the same dynamics governs the larger meteoroids as govern the comet. The difference in radiation forcing between the stream meteoroids and the comet is small, and if collisions are not important, then the uncertainties in the elements of the meteoroids will also grow identically and independently in the same manner as the comet's elements.  Our numerical simulations (which do not include non-gravitational forces for the comet) verify that the dynamics of the comet are not significantly distinguishable from those of the meteoroids, so that it is just one element of the cloud.  Thus the comet and each meteoroid in the stream can be viewed as a variant of the initial conditions.  Growth of the dispersion of the cloud is diagnostic of the growth in uncertainty in the state of the comet, and vice versa.

\begin{figure}[p!]
\begin{center}
\includegraphics[width=7in]{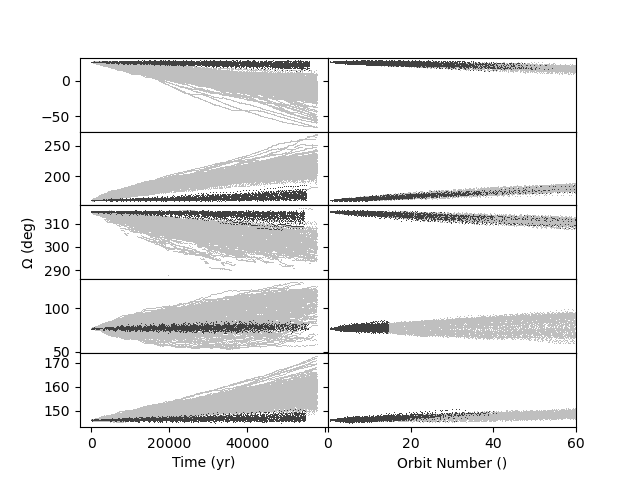}
\caption[]{
    \label{Fig20_Omega_vtandn}
The nodal dispersion as a function of time and orbit number. The values of $\Omega$ at Earth crossing are shown versus time (left) and orbit number (right) for five showers, each of which was simulated with initial periods of 250 and 4000 years.  The values from the short period simulations are shown in light gray, and for the long period in black.  The short period showers are abridged in the right-hand plot, because they go through hundreds of orbits but only 60 are shown.  The long period showers never experience more than 60 orbits.
The changes in nodal progression and dispersion occur at different rates versus time, but the simulations behave statistically identically versus orbit number.
}
\end{center}
\end{figure}

The model supposes that a cloud's dispersion is determined entirely by the impulses at perihelia.  This implies that the number of orbits is the most important factor in determining the evolution.
Figure~\ref{Fig20_Omega_vtandn} demonstrates that this is indeed so via plots of the simulated values of $\Omega$ versus time (left) and orbit number (right) for the five of our showers for which we have long term simulations with initial periods of 250 y and 4000 y.  We chose $\Omega$ to plot because the effects of nodal progression make for an easily identifiable systematic change.  The plotted values are from clouds of 300 particles, followed over 50-60kyr, and the figure shows the value of $\Omega$ for each particle at each Earth-crossing.  Showers having 250 y periods are shown in light gray, and those having 4000 y periods in black. The nodal progression is much slower for the longer period simulations, but when plotted versus orbit number, the rate of nodal progression and the dispersion growth are seen to behave identically in the short and long period simulations.  Indeed, we verified that the relations in Equations~\ref{eqn:sigma_elements1}-~\ref{eqn:sigma_elements3} appear to be the same for 4000y and 250y comets.

\begin{figure}[p!]
\begin{center}
\includegraphics[width=7in]{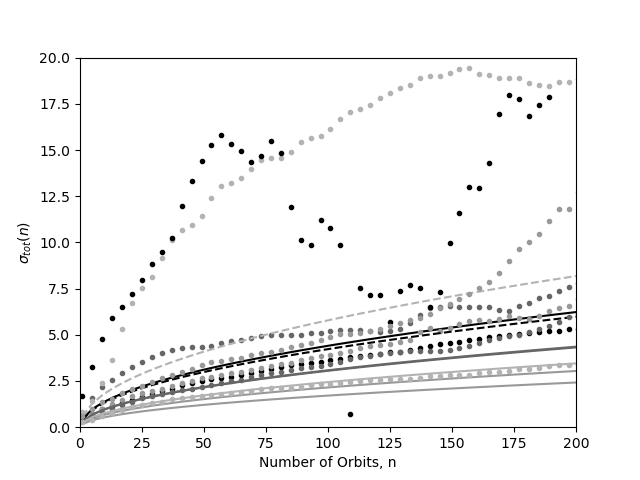}
\caption[]{
    \label{Fig21_sqrtn_growth}
Comparing numerical models of stream dispersion to results predicted by Equation~\ref{eqn:sigsquared_formula}. The growth of the total dispersion (Equation~\ref{eqn:sigma_rss}) from simulations (dots) is plotted versus orbit number along with the $n^{1/2}$ growth predicted from Equation~\ref{eqn:sigsquared_formula} (lines), for eight comets simulated with 250 y periods.  The reason for the anomolous behavior of two of the comets is explained in the text, and the model curves for those are shown as dotted instead of solid lines.   
}
\end{center}
\end{figure}

Figure~\ref{Fig21_sqrtn_growth} compares the growth of the total dispersion versus orbit number derived from integrations with model predictions, for the eight of our comets for which we had long simulations at 250 y.  The tabulated rates of dispersion growth are shown as dots, with different shades for each shower.  The model dispersions are the plotted curves in corresponding shades, calculated from Equation~\ref{eqn:var_over_orbit} with values of $\sigma^2_{\Delta j}$ taken from the relations in Equations~\ref{eqn:sigma_elements1}-\ref{eqn:sigma_elements3}.  For all but two of the showers the dispersion grows roughly in line with the $n^{1/2}$ behavior, with the same general magnitude as predicted a priori from the orbital geometries.  The two exceptional showers are \#16, which has a low value of $q\sim 0.27$ and for which the simulation split into distinct streams, and shower \# 1191 which drifted away from Earth crossing and consequently suffers from low number statistics.  For all the others, the growth is generally slightly larger than would be predicted from the model, presumably because we did not include planetary encounters.

\subsection{Application to observations of dispersion growth}

When observing meteor showers, we do not have the option of tabulating meteoroid crossings by the number of orbits they have completed.  We observe only samples of the stream population at particular times, some members of which have undergone many orbits, and some of which have undergone only a few.  Meteoroids that have been in orbits with small periods will be observed after having crossed a larger number of perihelia than those that have drifted to high period orbits.  
We find that the distribution of the number of orbits completed on arrival at Earth in a meteoroid stream at any particular time, which we will refer to as $N_t(n)$, is well characterized at each time by a Maxwellian distribution, and that the mean and variance of the distribution seems to grow somewhat similarly for all showers.

With knowledge of $N_t(n)$, we can compute the statistics of the dispersions observed at particular times by integrating over the distribution of orbit numbers:
\begin{eqnarray}
  {\rm E}\left[\mathbf{S}_j(t)\right] & = & S_{j,0} + 
            \sum_{i=1}^\infty \left(N_t(n) n \Delta_{\tau}{\rm S}_j \right)\\
  {\rm V}\left[\mathbf{S}_j(t)\right] & = & 
          \sum_{i=1}^{\infty} 
                 N_t(i) {\rm V}\left[\mathbf{S}_j(i))\right] +  
          \sum_{i=1}^\infty 
                  N_t(i)\left({\rm E}\left[\mathbf{S}_j(i)\right] -
                                {\rm E}\left[\mathbf{S}_j(t)\right]\right)^2
\label{eqn:Eq_WeightedSum}
\end{eqnarray}
where the sums are over orbit numbers, $i$, $N_t(i)$ is the frequency of meteoroids at time $t$ having completed exactly $i$ orbits, and ${\rm V}[\mathbf{S}_j(i)]$ is the variance after $i$ orbits.  The very last term in Equation~\ref{eqn:Eq_WeightedSum} contains the expectation of ${\rm S}_j$ for each orbit number minus the expectation of ${\rm S}_j(t)$ over all orbits at time, $t$.

The two summations on the right hand side of Equation~\ref{eqn:Eq_WeightedSum} separate the total variance computed at each time into a variance within groups and variance between groups, also known in ANOVA analysis as SSE and SST (e.g.,\cite{SchaefferMW}).
The first term expresses the intrinsic dispersion after $n$ orbits due to the randomization introduced by the Sun.  If the process were stationary, so that the means of $\mathbf{S}_j$ were constant, then this would be the only term.  The second term accounts for the fact that when computing the variance of a sample of meteoroids with varying numbers of orbits, there is a contribution from the drift of the mean with orbit number.  

Using expressions for the expectation and variance of the orbital elements as functions of orbit from Equations~\ref{eqn:expec_over_orbit} and~\ref{eqn:var_over_orbit}, it follows after some algebra that the expectations and variances as a function of time are
\begin{eqnarray}
    {\rm E}\left[S_j(t)\right] & =  & S_0 + \overline{N_t(n)} (\Delta_\tau\mathbf{S}_j)  
    \label{eqn:sigsquared_formula1}\\
    {\rm V}\left[S_j(t)\right] & = & \overline{N_t(n)}(\sigma^2_{\Delta S_j}) + 
                       {\rm V}\left[N_t(n)\right]\Delta^2_{\tau}\mathbf{S}_{j},
\label{eqn:sigsquared_formula} 
\end{eqnarray}
where $\overline{N_t(n)}$ and ${\rm V}\left[N_t(n)\right]$ are the mean and variance, respectively of $N_t(n)$ at time $t$, taken with respect to orbit number.  The expected value of $\mathbf{S}_j$ is merely the accumulation of $n$ steps of torque-induced precession, whereas the variances within and between orbit numbers are proportional to the mean and variance of $N_t(n)$, respectively.

\begin{figure}[p!]
\begin{center}
\includegraphics[width=7in]{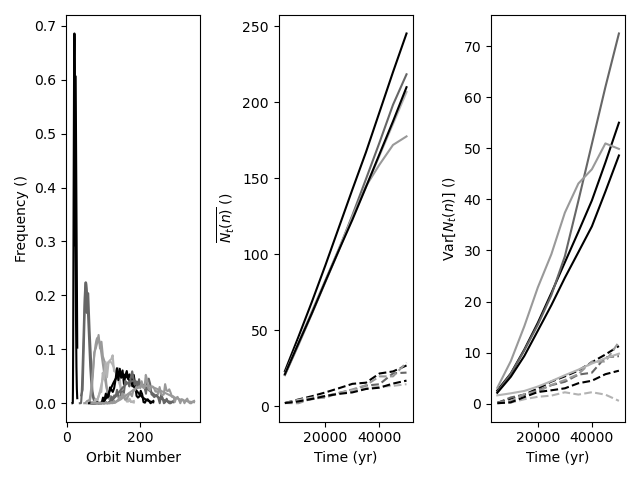}
\caption[]{
    \label{Fig22_Maxwellians}
Maxwellian distribution of the number of completed orbits on encounter with Earth at different times (right), and the mean and variance of the number of orbits of meteoroids encountered at Earth over time. Solutions for orbits with initial orbital period 250y (solid lines) and 4000y (dashed) lines are shown.
}
\end{center}
\end{figure}

The behavior of $N_t(n)$ is demonstrated in Figure~\ref{Fig22_Maxwellians}.  The left-hand panel shows an example of Maxwellian fits to $N_t(n)$ for shower \#839 calculated at every 8000 years from a simulation that followed 300 particles over 60kyr.  All particles in the simulation crossing the node within 500 years of the selected time, and within 0.05 AU of Earth's orbit, were used to construct the histograms fitted by the distributions.  Generally all the fits for all showers and times are of this quality, becoming noisier as the stream loses particles and the distribution widens.

The middle and right-hand panels show the time-wise growth of the mean and variance, respectively, for the five of our showers for which we had 60kyr simulations with periods of 250 and 4000 years.
The means grow nearly linearly versus orbit, and the variances grow slowly at first but then grow faster than the means as the distribution widens and the variance of the means between different orbits (variance between groups) becomes appreciable.  
The mean and variance for the long period simulations grow more slowly than those of the shorter period simulations, roughly by the ratio of orbital periods.

\begin{figure}[p!]
\begin{center}
\includegraphics[width=7in]{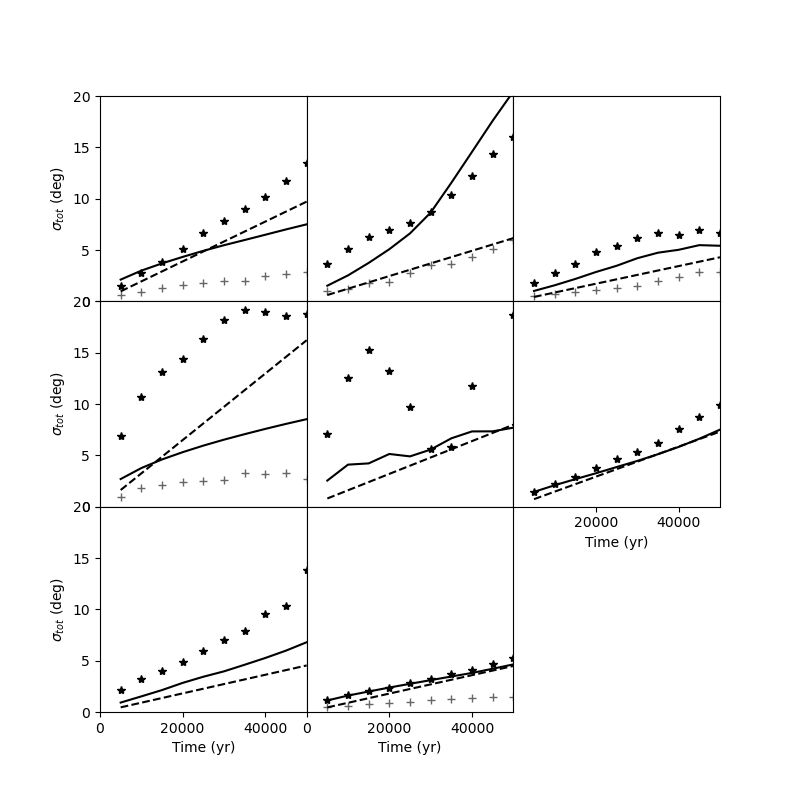}
\caption[]{
    \label{Fig23_Disps_v_time}
Increase of meteoroid stream dispersion with time from the modeling results of eight meteor showers with assumed initial orbital periods of 250y (*) and 4000y (+). The dashed line shows the predicted behavior from Equation~\ref{eqn:sigma_age}. The solid line is that calculated from the Sun Close-Encounter model. Note how the simple relation characterizes the behavior well in all cases, with the dashed line falling in between the 250 and 4000y results where those results are most different.
}
\end{center}
\end{figure}

Figure~\ref{Fig23_Disps_v_time} shows the dispersion growth from simulations along with model predictions for our 8 sample showers using the above apparatus.  The dispersion calculated directly from the simulation, $\sigma_{\rm tot}$, is evaluated every 10kyr and shown in stars for the simulations with 250 year initial periods, and crosses for the 5 showers for which we had 4000 year initial periods.  Because the evolution is determined by orbit numbers, the dispersion growth for the longer period showers is slower.  When looking at the dispersion of the whole cloud we do not include the attack angle, but use it when we look at the dispersion of the particles intersecting Earth's orbit.

The model dispersions are calculated using Equation~\ref{eqn:sigsquared_formula} 
with the $\sigma_{0,\Delta S_j}$ estimated from Equations~\ref{eqn:sigma_elements1}-~\ref{eqn:sigma_elements3} and a fitted Maxwellian. \citet{Jenniskens2023}'s linear model from Equation~\ref{eqn:sigma_rss} is shown as a dashed line. 
Showers \#16 and \#1191, the left-most two panels in the middle row, are the showers that showed anomalous growth in Figure~\ref{Fig21_sqrtn_growth} and also are not well-fit by the models in this figure.

The secular evolution and time-wise dispersion growth is dominated by the rate at which the number of orbits grows over time.  This is basically determined by the distribution of orbital periods, which is not known. The model allows an evaluation of the rate of growth for different period showers, and the rates are not extremely sensitive to the values of $\sigma_{0,j}$ that are chosen, which appear to be always approximately $\sigma_{\rm tot}\approx 0.5^\circ.$  
Without the contributions to the dispersion from the variance of $N_t(n),$ the growth of the variances in time would be determined by $\overline{N_t(n)}$, and so the dispersions would go as $\overline{N_t(n)}^{1/2}.$  The more linear rate of growth seen here appears to be due to the last term of Equation~\ref{eqn:Eq_WeightedSum} taking over, with the variance-between-groups contributing more and more dispersion as observed number of orbits increases.

\citet{Jenniskens2023}'s simple linear fit from Equation~\ref{eqn:sigma_rss} (shown as a dashed line in Figure~\ref{Fig23_Disps_v_time}) does not include information about the average orbital period within a stream, as it was derived trying to avoid information that is not observed.  The difference in slopes between short and long period showers reflects the simple relation that ignoring the distribution of the periods within a stream, the age would be the number of orbits times the orbital period.  We find that \citet{Jenniskens2023}'s relation generally falls in between the solutions for orbital periods of 250 and 4000 y, respectively. The result is typically within a factor of two of that found in numerical modeling. With one or two exceptions where the 250 and 4000y solutions are most divergent, the results from the Sun Close-Encounter model agree well with Equation~\ref{eqn:sigma_rss}. No clear systematic difference was found that requires a re-calibration of the relation. 

We postulate that the reason that \citet{Jenniskens2023} gets distinct slopes for general LPCs versus shorter-orbit Mellish-type showers is due to a difference in orbital periods, manifest in the above model as differences in $N_t(n)$. Our numerical simulations indicate that comets with starting periods of 250 to 4000 years almost always exhibit a drop of mean orbital period of 200 to 250 years over time spans of 30-50kyr which would correspond to the derived slope of $1/(12,000 q).$ This drop is at least partially due to counting bias, in that the meteoroids with longer orbital periods do not contribute as much to the average period as those which evolve to shorter periods, because those meteoroids are undersampled by finite-time observations.  

In contrast, the meteoroids in the Mellish-type comet showers are thought to get captured into shorter orbits on the order of 10 to 160 years. At any given time, the bulk of the stream meteoroids would have undergone approximately four times the number of orbits than those in a similarly aged LPC.  If the growth goes as $n^{1/2}$, a factor of four difference in the number of orbits crossed by the Mellish-type comets would result in a dispersion growth rate twice as high, explaining the derived slope of $1/(6000\,q)$ for the Mellish showers versus that $1/(12,000\,q)$ for the LPCs.

\subsection{Additional Considerations}

The model presented here addresses the growth in dispersion of a cloud of ejecta only from a single perihelion, whereas annual observed meteor showers are expected to contain meteoroids that were ejected during at least several perihelion crossing.  A simple line of reasoning suggests that while the total number of particles in the multi-sourced stream will grow at a rate determined by the balance between injection and loss rates of meteors over many orbits, the rate of growth of dispersion is governed simply by perturbations, and should not sum over streams.  

The effect of the initial ejection conditions seems to be lost after a few orbits, but we have not studied that in the present work. An initial rapid growth in dispersion is present in simulations that lasts for a few orbits.  Our numerical simulations for streams assign initial velocities for ejecta that are radially outward from the lit face of the comet, with magnitudes of up to 10m/s, and the initial dispersion could relate to the details of this.  The subsequent dispersion growth appears to be dominated by the variance in perturbations at perihelion crossings and because different parts of dust trails catch up on each other.  

Our numerical simulations indicate that all of the elements, as well as the distribution of orbits, $N_t(n),$ are well-fit at all times by Maxwellian distributions.  This behavior is expected for random walks in three or more dimensions.  In simulations, the mean of the distribution, $\overline{N_t(n)},$ increases linearly in time and the variance broadens.  Application to general observational data requires a theoretical understanding of this growth and spreading, but here we have merely checked the approach by using numerical simulations to evaluate $N_t(n)$ together with the dispersion growth.  

In principle, planetary torque can be added as part of the terms $\Delta_{pl}$ and $\Delta_\tau$ that appear in Equation~\ref{eqn:markovstate}. The planets appear to exert their torque at a ``Goldilocks zone'' at approximately 10AU, where the force is strong and lever arm is big, hence their effects are separable from perturbations due to the Sun, which occur within 5AU.  A method like Gaussian averaging could be used to predict the expected torque from a 'disk' at each crossing.  The planetary perturbations treated thus would have a high variance, being less frequent but with the possibility of being quite large. Because they affect individual orbits, it is not clear how they would add to the dispersion of meteoroids observed at Earth. In some cases they would simply lead to loss of meteoroids from the stream, adding to the meteor sporadic background.  In others, as with the repeated encounters with Jupiter in lowly inclined showers \#16 and \#206 studied here, a systematic change to the growth rate might occur.

\section{Conclusions}

The numerical simulations of comets and meteoroid streams presented here indicate that LPCs and their associated meteoroids spend most of their existence in stable, barycentric orbits, which are perturbed into new orbits at each perihelion crossing.  The nature of this perturbation is that the comet experiences a 'close encounter' with the Sun, which pulls it from its barycentric orbit into an elliptical heliocentric orbit for a few hundred days on either side of perihelion, when the comet is inside the orbit of Jupiter.  Although planetary torques and close encounters are present, the perturbations of the orbital elements from one barycentric orbit to the next are most often dominated by the state of the Sun relative to the comet near perihelion, as they both orbit SSB.

Energy and momentum exchange with the Sun perturbs $a$ and $q,$ while the perturbation to the angular momentum vector alters the angular elements, $\Omega, i$ and $\omega.$  The period and perihelion distance change by gravitational boost or braking, which depend upon the barycentric velocity direction of the Sun relative to the comet's eccentricity vector during encounter. The angular elements respond to the perturbation in angular momentum resulting from the fact that the Sun is not generally in the incoming barycentric orbital plane when the comet enters its sphere of influence, nor is  the barycenter is in the heliocentric orbital plane when the comet leaves the Sun's sphere of influence.  This necessitates changes in the direction of the angular momentum when the Sun takes and then releases control of the comet, which in turn determine the changes to the orbital elements $\Omega, i$ and $\omega.$ These changes are dominated by the perpendicular distance of the Sun from the incoming barycentric orbital plane and its velocity.  

The long term evolution of orbits can be formulated as a Markov random walk in the orbital elements, in which a step is taken at each perihelion, driven by a random sampling of the Sun's state at encounter.  The variances in the distributions of perturbations, or kicks, to the angular elements at each perihelion result from the statistics of the Sun's orbit about barycenter, such that their probability distributions can be calculated {\it a priori} for any given orbital geometry. The dispersion growth depends on the perihelion distance as $q^{-\sqrt{2}/2},$ due to the way in which $q$ alone determines the shape of LPC orbits within 5AU.  The root-sum-square of the kicks, $\sigma_{tot}$, has a mean of about $0.4^\circ$ for the twelve comets studied in detail here.  If the comet orbit is in a configuration where the inclination and argument of perihelion are stationary, then the dispersion grows as the square root of the number of orbits. 

Two strong conclusions of this work are that the statistical uncertainty in the elements of any stream meteoroid is determined primarily by the number of orbits it has undergone, and that the dispersion of a stream can be found by summing over the uncertainties in locations of the individual meteoroids within it.  From these assumptions it follows that in order to estimate the dispersion of a stream at any time, one must make use of the distribution, $N_t(n),$ of orbits that are observed at any time.  Any analysis of dispersions evaluated at fixed times requires some summation over an estimate of the distribution of orbit numbers expected at those times.  If the solar forcing terms are constant over time and can be taken outside the summations, then the dispersion at any time depends on the mean and variance of the distribution $N_t(n)$.  Our simulations indicate that $N_t(n)$, as well as the distributions of each of the elements individually, are well-fit by a Maxwellians.  This is consistent with the elements following a random walk in three or more dimensions.  
 
The Sun Close-Encounter model of LPCs and their meteoroid stream evolution, in which they are described as undergoing sequences of barycentric orbits punctuated by solar close encounters inside the orbit of Jupiter, seems to capture the main physical effects governing their long-time evolution. LPCs are unique in that they are not appreciably influenced by the constant ebbs and flows of all solar system bodies but see only a snapshot of the inner solar system in passing through perihelion, and that snapshot is summarized to first order by the state of the Sun itself. General torque from the planets has maximum effect near 10AU, and can be modeled with Gaussian averaging over the inner planets and Jupiter. Planetary close encounters occur sporadically, but they typically do not dominate the rate of dispersion. 

The Sun Close-Encounter model establishes a physical basis for understanding the growth of dispersion in streams, and provides a framework for the statistical study of these streams using stochastic Markov random walks.

\section{Acknowledgments}
The authors thank J. Vaubaillon of I.M.C.C.E., Paris Observatory, for making the PINTEM software package available to us. The work was supported by NASA YORPD grant 80NSSC22K1467.  The authors thank Dr. A. Sekhar for valuable corrections, and an anonymous reviewer for pointing out the work of \citet{Carusi1987} and other suggestions making this a better paper.




\printcredits

\bibliographystyle{cas-model2-names}

\bibliography{Citations}

\begin{thebibliography}{16}
\expandafter\ifx\csname natexlab\endcsname\relax\def\natexlab#1{#1}\fi
\providecommand{\url}[1]{\texttt{#1}}
\providecommand{\href}[2]{#2}
\providecommand{\path}[1]{#1}
\providecommand{\DOIprefix}{doi:}
\providecommand{\ArXivprefix}{arXiv:}
\providecommand{\URLprefix}{URL: }
\providecommand{\Pubmedprefix}{pmid:}
\providecommand{\doi}[1]{\href{http://dx.doi.org/#1}{\path{#1}}}
\providecommand{\Pubmed}[1]{\href{pmid:#1}{\path{#1}}}
\providecommand{\bibinfo}[2]{#2}
\ifx\xfnm\relax \def\xfnm[#1]{\unskip,\space#1}\fi
\bibitem[{Acton(1996)}]{Acton1996}
\bibinfo{author}{Acton, C.H.}, \bibinfo{year}{1996}.
\newblock \bibinfo{title}{Ancillary data services of nasa's navigation and
  ancillary information facility}.
\newblock \bibinfo{journal}{Planetary and Space Science} \bibinfo{volume}{44},
  \bibinfo{pages}{65--70}.
\bibitem[{Carusi et~al.(1987)Carusi, Kresak, Perrozi and
  Valsecchi}]{Carusi1987}
\bibinfo{author}{Carusi, A.}, \bibinfo{author}{Kresak, L.},
  \bibinfo{author}{Perrozi, E.}, \bibinfo{author}{Valsecchi, G.},
  \bibinfo{year}{1987}.
\newblock \bibinfo{title}{High-order librations of halley-type comets}.
\newblock \bibinfo{journal}{Astronomy and Astrophysics} \bibinfo{volume}{187},
  \bibinfo{pages}{899--905}.
\bibitem[{Chambers(1997)}]{Chambers1997}
\bibinfo{author}{Chambers, J.E.}, \bibinfo{year}{1997}.
\newblock \bibinfo{title}{Why halley-types resonate but long-period comets
  don't: A dynamical distinction between short- and long-period comets}.
\newblock \bibinfo{journal}{Icarus} \bibinfo{volume}{125},
  \bibinfo{pages}{32--38}.
\bibitem[{Everhart(1969)}]{Everhart1969}
\bibinfo{author}{Everhart, E.}, \bibinfo{year}{1969}.
\newblock \bibinfo{title}{Close encounters of comets and planets}.
\newblock \bibinfo{journal}{Astronomical Journal} \bibinfo{volume}{74},
  \bibinfo{pages}{735--750}.
\bibitem[{Gastineau et~al.(2024)Gastineau, Laskar, Fienga and
  Manche}]{Gastneau2024}
\bibinfo{author}{Gastineau, M.}, \bibinfo{author}{Laskar, J.},
  \bibinfo{author}{Fienga, A.}, \bibinfo{author}{Manche, H.},
  \bibinfo{year}{2024}.
\newblock \bibinfo{title}{Calceph - c language. release 3.5.5}.
\newblock
  \bibinfo{howpublished}{\url{https://www.imcce.fr/content/medias/recherche/equipes/asd/calceph/calceph_c.pdf}}.
\newblock \bibinfo{note}{[Accessed 22-May-2024]}.
\bibitem[{Hajduková and Neslusan(2019)}]{Hajdukova2019}
\bibinfo{author}{Hajduková, M.}, \bibinfo{author}{Neslusan, L.},
  \bibinfo{year}{2019}.
\newblock \bibinfo{title}{Modeling of the meteoroid stream of comet c/1975 t2
  and $\lambda$-ursae majorids}.
\newblock \bibinfo{journal}{Astronomy and Astrophysics} \bibinfo{volume}{627},
  \bibinfo{pages}{id.A73, 8 pp}.
\bibitem[{Hajduková and Neslusan(2021a)}]{Hajdukova2021a}
\bibinfo{author}{Hajduková, M.}, \bibinfo{author}{Neslusan, L.},
  \bibinfo{year}{2021}a.
\newblock \bibinfo{title}{Modeling the meteoroid streams of comet c/1861 g1
  (thatcher), lyrids}.
\newblock \bibinfo{journal}{Planetary and Space Science} \bibinfo{volume}{203},
  \bibinfo{pages}{id.105246}.
\bibitem[{Hajduková and Neslusan(2021b)}]{Hajdukova2021b}
\bibinfo{author}{Hajduková, M.}, \bibinfo{author}{Neslusan, L.},
  \bibinfo{year}{2021}b.
\newblock \bibinfo{title}{Modeling the meteoroid streams of comets c/1894 g1
  (gale) and c/1936 o1 (kaho-kozik-lis)}.
\newblock \bibinfo{journal}{Planetary and Space Science} \bibinfo{volume}{195},
  \bibinfo{pages}{id.105152}.
\bibitem[{Jenniskens(1997)}]{Jenniskens1997}
\bibinfo{author}{Jenniskens, P.}, \bibinfo{year}{1997}.
\newblock \bibinfo{title}{Meteor stream activity. iv. meteor outbursts and the
  reflex motion of the sun}.
\newblock \bibinfo{journal}{Astronomy and Astrophysics} \bibinfo{volume}{317},
  \bibinfo{pages}{953--961}.
\bibitem[{Jenniskens(2023)}]{Jenniskens2023}
\bibinfo{author}{Jenniskens, P.}, \bibinfo{year}{2023}.
\newblock \bibinfo{title}{Atlas of Earth's Meteor Showers}.
\newblock \bibinfo{publisher}{Elsevier Science}, \bibinfo{address}{Amsterdam,
  Netherlands}.
\bibitem[{Jenniskens et~al.(2021)Jenniskens, Lauretta, Towner, Heathcote,
  Jehin, Hanke, Cooper, Baggaley, Howell, Johannink, Breukers, Odeh, Moskovitz,
  Juneau, Beck, De~Cicco, Samuels, Rau, Albers and Gural}]{Jenniskens2021}
\bibinfo{author}{Jenniskens, P.}, \bibinfo{author}{Lauretta, D.S.},
  \bibinfo{author}{Towner, M.C.}, \bibinfo{author}{Heathcote, S.},
  \bibinfo{author}{Jehin, E.}, \bibinfo{author}{Hanke, T.},
  \bibinfo{author}{Cooper, T.}, \bibinfo{author}{Baggaley, J.W.},
  \bibinfo{author}{Howell, J.A.}, \bibinfo{author}{Johannink, C.},
  \bibinfo{author}{Breukers, M.}, \bibinfo{author}{Odeh, M.},
  \bibinfo{author}{Moskovitz, N.}, \bibinfo{author}{Juneau, L.},
  \bibinfo{author}{Beck, T.}, \bibinfo{author}{De~Cicco, M.},
  \bibinfo{author}{Samuels, D.}, \bibinfo{author}{Rau, S.},
  \bibinfo{author}{Albers, J.}, \bibinfo{author}{Gural, P.S.},
  \bibinfo{year}{2021}.
\newblock \bibinfo{title}{Meteor showers from known long-period comets}.
\newblock \bibinfo{journal}{Icarus} \bibinfo{volume}{365},
  \bibinfo{pages}{id.114469}.
\bibitem[{Lyytinen and Jenniskens(2003)}]{Lyytinen2003}
\bibinfo{author}{Lyytinen, E.}, \bibinfo{author}{Jenniskens, P.},
  \bibinfo{year}{2003}.
\newblock \bibinfo{title}{Meteor outbursts from long-period comet dust trails}.
\newblock \bibinfo{journal}{Icarus} \bibinfo{volume}{162},
  \bibinfo{pages}{443–452}.
\bibitem[{Neslusan and Hajduková(2021)}]{Neslusan21}
\bibinfo{author}{Neslusan, L.}, \bibinfo{author}{Hajduková, M.},
  \bibinfo{year}{2021}.
\newblock \bibinfo{title}{Meteoroid stream of comet c/1961 t1 (seki) and its
  relation to the december-virginids and -sagittariids}.
\newblock \bibinfo{journal}{The Astronomical Journal} \bibinfo{volume}{162},
  \bibinfo{pages}{id.20, 9pp}.
\bibitem[{Pilorz et~al.(2023)Pilorz, Jenniskens and Vaubaillon}]{Pilorz2023}
\bibinfo{author}{Pilorz, S.}, \bibinfo{author}{Jenniskens, P.},
  \bibinfo{author}{Vaubaillon, J.}, \bibinfo{year}{2023}.
\newblock \bibinfo{title}{Age-dependent orbital dispersion growth of long
  period comet meteor showers}, in: \bibinfo{booktitle}{Asteroids, Comets,
  Meteors Conference 2023}, p. \bibinfo{pages}{2484}.
\bibitem[{Schaeffer et~al.(2008)Schaeffer, Mendenhall and
  Wackerly}]{SchaefferMW}
\bibinfo{author}{Schaeffer, R.}, \bibinfo{author}{Mendenhall, M.},
  \bibinfo{author}{Wackerly, D.}, \bibinfo{year}{2008}.
\newblock \bibinfo{title}{Mathematical Statistics with Applications, 7th Ed.}
\newblock \bibinfo{publisher}{Brooks/Cole CENGAGE}.
\bibitem[{{Vaubaillon} et~al.(2005){Vaubaillon}, {Colas} and
  {Jorda}}]{Vaubaillon2005}
\bibinfo{author}{{Vaubaillon}, J.}, \bibinfo{author}{{Colas}, F.},
  \bibinfo{author}{{Jorda}, L.}, \bibinfo{year}{2005}.
\newblock \bibinfo{title}{{A new method to predict meteor showers. I.
  Description of the model}}.
\newblock \bibinfo{journal}{Astronomy and Astrophysics} \bibinfo{volume}{439},
  \bibinfo{pages}{751--760}.

\end{thebibliography}

\pagebreak

\begin{table}[]
\caption{List of meteor showers studied in numerical modeling. Some of the proposed parent bodies are uncertain \citep{Jenniskens2023}.  Showers shown in bold are those studied in detail here.}{}
\label{table1}

\begin{tabular}{ c c c c c c c }
\toprule
Shower & Code & q     & i   & Peri & Node  & Parent Body  \\ \hline
6    & LYR & 0.920      & 79.4  & 214.2    & 32.3 & C/1861 G1    \\ 
\textbf{16}  & \textbf{HYD} & 0.257      & 128.8  & 119.3    & 76.6 & C/2023 P1    \\ 
\textbf{16a}   &\textbf{HYD} & 0.257      & 30.0 & 119.3    & 76.6 &  -.-   \\ 
\textbf{16b}    &\textbf{HYD} & 0.257      & 60.0  & 119.3    & 76.6 & -.-   \\ 
130    & DME & 0.963      & 59.6  & 353.0    & 179.7 & -.-    \\ 
145    & ELY & 1.000      & 74.4  & 191.3    & 50.0 & C/1981 H1   \\ 
\textbf{206}    &\textbf{AUR} & 0.677      & 148.0  & 109.4    & 158.9 & C/1911 N1    \\ 
331    & AHY & 0.314      & 61.5  & 112.9    & 97.1 & -.-    \\ 
388    & CTA & 0.090      & 16.6 & 325.8    & 221.9 & C/1953 X1        \\ 
410    & DPI & 0.920      & 178.3  & 143.8    & 90.8 & C/1864 N1    \\ 
\textbf{427}    &\textbf{FED} & 0.971      & 55.4  & 194.3    & 315.2 & -.-    \\ 
458    & JEC & 0.920      & 95.7  & 215.8    & 82.4 & -.-    \\ 
510    & JRC & 1.007      & 88.5  & 190.9    & 84.2 & C/2003 T4    \\ 
517    & ALO & 0.303      & 110.8  & 293.7    & 16.8 & -.-    \\ 
\textbf{519}   & \textbf{BAQ} & 0.919      & 156.0  & 144.9    & 41.8 & -.-    \\ 
569    & OHY & 0.680      & 114.3  & 69.4    & 129.5 & C/770 K1    \\ 
571    & TSB & 0.497      & 83.1  & 270.3    & 343.3 & -.-    \\ 
\textbf{647}    &\textbf{BCO} & 0.640      & 22.1  & 250.5    & 7.3 & -.-    \\ 
681    & OAQ & 0.363      & 161.9  & 287.8    & 91.9 & -.-    \\ 
752    & AAC & 0.794      & 167.3  & 125.4    & 17.66 & C/1917 H1    \\ 
792    & MBE & 0.167      & 73.9  & 48.1    & 355.5 & -.-    \\ 
\textbf{822}    &\textbf{NUT} & 0.562      & 135.9  & 275.2    & 318.1 & C/1797 P1    \\ 
828    & TPG & 0.247      & 49.2  & 59.3    & 5.4 & -.-    \\ 
\textbf{839}    &\textbf{PSR} & 0.429      & 69.0  & 278.1    & 25.0 & -.-    \\ 
853    & ZPA & 0.993      & 99.6  & 353.6    & 181.4 & -.-    \\ 
1045    & SUT & 0.657      & 175.2  & 252.4    & 179.3 & C/2005 T4    \\ 
\textbf{1047}    &\textbf{GCR} & 0.929      & 102.8  & 29.3    & 145.9 & -.-    \\ 
\textbf{1055}    &\textbf{TVL} & 0.570      & 88.1  & 97.7    & 22.0 & -.-    \\ 
1056    & AZC & 0.650      & 80.3  & 106.9    & 21.5 & -.-    \\ 
1146    & DHO & 0.907      & 74.1  & 37.4    & 348.5 & -.-    \\ 
\textbf{1191}    &\textbf{EIV} & 0.522      & 165.4  & 267.5    & 321.4 & -.-    \\ \hline
\bottomrule
\end{tabular}

\end{table}

\begin{table}[]
\caption{Overview of parameters that influence the kick and precession of the orbital elements.}{}
\label{table1}

\begin{tabular}{c c c }
\toprule
Element & Random Kick & Precession  \\ \hline
a & Energy exchange in Sun close encounter & -.- \\
e & Energy exchange in Sun close encounter & -.- \\
q & Energy and impulse & Torque \\ 
i & Impulse from Sun not in orbital plane & Torque from Sun and planets \\
$\omega$ & Impulse from Sun not in orbital plane & Torque from Sun and planets  \\
$\Omega$ & Impulse from Sun not in orbital plane & Torque from Sun and planets \\
\\ \hline
\bottomrule
\end{tabular}

\end{table}

\pagebreak


\end{document}